\documentclass[12pt, leqno]{article}
\usepackage{amsmath,caption,setspace,multirow}
\usepackage[top=1.12in, bottom=1.12in, left=1.1in, right=1.1in]{geometry}

\usepackage[round]{natbib}
\usepackage{color,soul}

\DeclareCaptionStyle{italic}[justification=centering]{labelfont={bf},textfont={it},labelsep=colon}
\usepackage{graphicx,psfrag,epsf,textcomp,epstopdf}
\usepackage{enumerate, dsfont}
\usepackage{natbib}
\usepackage{url,xcolor}
\usepackage{booktabs, subfig, bm, paralist,mathpazo,tikz,todonotes,longtable,microtype,algorithm, bbm}

\usepackage[pdftex,colorlinks=true]{hyperref}
\definecolor{darkblue}{rgb}{0,0,.6}
\hypersetup{citecolor=darkblue,linkcolor=darkblue,urlcolor=darkblue}
\definecolor{DarkRed}{rgb}{.7,0,.4}

\newcommand\Tau{\mathcal{T}}
\usepackage{comment}

\newcommand{\blind}{0}

\newcommand{\X}{\mathcal{X}}
\newcommand{\Y}{\mathcal{Y}}

\graphicspath{{plots/}}

\newsavebox\CBox

 \newtheorem{@definition}{\sc Definition}[section]

  \renewcommand\X{\mathcal{X}}

\begin{document}

\def\spacingset#1{\renewcommand{\baselinestretch}{#1}\small\normalsize} \spacingset{1}

\if0\blind
{
  \title{\bf Functional linear models for interval-valued data}
    \author{
Ufuk Beyaztas \\
Department of Statistics \\
Bartin University \\
\\
Han Lin Shang \\
Research School of Finance, Actuarial Studies and Statistics \\
Australian National University
\\
\\
Abdel-Salam G. Abdel-Salam\\
Department of Mathematics, Statistics and Physics \\
Qatar University 
 }
  \maketitle
} \fi

\if1\blind
{
  \bigskip
  \bigskip
  \bigskip
  \begin{center}
    {\LARGE\bf Functional linear models for interval-valued data}
\end{center}
  \medskip
} \fi

\maketitle

\begin{abstract}
Aggregation of large databases in a specific format is a frequently used process to make the data easily manageable. Interval-valued data is one of the data types that is generated by such an aggregation process. Using traditional methods to analyze interval-valued data results in loss of information, and thus, several interval-valued data models have been proposed to gather reliable information from such data types. On the other hand, recent technological developments have led to high dimensional and complex data in many application areas, which may not be analyzed by traditional techniques. Functional data analysis is one of the most commonly used techniques to analyze such complex datasets. While the functional extensions of much traditional statistical techniques are available, the functional form of the interval-valued data has not been studied well. This paper introduces the functional forms of some well-known regression models that take interval-valued data. The proposed methods are based on the function-on-function regression model, where both the response and predictor/s are functional. Through several Monte Carlo simulations and empirical data analysis, the finite sample performance of the proposed methods is evaluated and compared with the state-of-the-art.
\end{abstract}


\newpage
\spacingset{1.56}

\section{Introduction}
Due to recent technological advances, the process of collecting data has become complicated, causing high dimensional and complex data structures. Symbolic data analysis is one of the commonly used methods in modeling such complex and large datasets, see \cite{billard2011} and \cite{noirhomme2011} for recent developments in symbolic data analysis. Contrary to single-valued observations in $p$-dimensional space where classical statistical methods work on, symbolic data may be in the form of hypercubes in $p$-dimensional space. There are many symbolic data types, for example, list, histogram, modal-valued, and interval-valued data. In this research, we restrict our attention to the interval-valued data only. The data expressed in an interval format (minimum and maximum values of the data) is called the interval-valued data. Such datasets are frequently encountered in daily life, for example, air and/or surface temperature, wind speed, energy production, blood pressure, and exchange rates. The main problem encountered during the modeling of the interval-valued data with classical statistical techniques is ``how the variability of observations within the range is involved in modeling?". Traditional methods analyze interval-valued data using its summary (i.e., mid-points), and this approach results in loss of information. Therefore, interval-valued data analysis techniques are needed to obtain more accurate information.

The early studies about the interval-valued data regression were conducted by \cite{billard2000}, who extended traditional statistical techniques to the interval-valued data. \cite{billard2002} extended several classical regression models to interval-valued data and they proposed a regression equation for fitting histogram-valued symbolic data; \cite{lima2004} proposed two interval-valued regression models using the mid-points and ranges of the interval values; \cite{alfonso2004} suggested a regression model for fitting taxonomic variables; \cite{billardbook} suggested several interval-valued regression models; \cite{billard2007} proposed a covariance function for the interval-valued data; \cite{maia2008} introduced an interval-valued regression model based on least absolute deviation; \cite{lima2008} suggested an interval-valued data regression model inspired by the work of \cite{lima2004}; \cite{lima2009} proposed a bivariate generalized linear model for interval-valued symbolic data; \cite{lima2010} presented a constrained interval-valued linear regression model; \cite{lima2011} introduced a bivariate symbolic regression model that considers the interval-valued variables as bivariate random vectors; \cite{ahn2012} proposed a resampling based interval-valued regression model; \cite{lim2016} suggested a nonparametric additive approach for analyzing interval-valued data that allows a nonlinear pattern; and \cite{limaNN} introduced a nonlinear regression model for interval-valued data and estimated the model parameters by several optimization algorithms.

The datasets repeatedly measured over discrete time points provide more information than those obtained from a single time point, where most of the interval-valued data regression studies have focused on in the pertinent literature. Available regression models may not be able to model such data type due to some common regression problems, such as high dimensionality, multicollinearity, and the high correlation between the sequential observations. On the other hand, functional data, which considers the data in the form of curves, can characterize this kind of data that is sampled over continuum measures. Functional Data Analysis (FDA) techniques reduce the problem of high dimensionality and focus on temporal dependence between the curves. It provides several advantages over the classical methods: for example, it is possible to look at the data as a whole; FDA is not affected by the missing data and the high correlation problem between the repeated measurements and minimizes the data noise by smoothing techniques. Consult \cite{ramsay2002, ramsay2005}, \cite{ferraty2006}, and \cite{KoRe} for more information about the FDA.

In this paper, we extend some well-known interval-valued data regression models to the functional data context. Functional regression models have become important analytical tools to explore the relationship between the response and predictor variables. In this context, several regression models have been proposed depending on whether response and/or predictor/s are scalar or functional; ($i$) functional response-scalar predictors; ($ii$) scalar response-functional predictors; and ($iii$) functional response-functional predictors. For the cases ($i$) and ($ii$), examples include \cite{cardot1999, cardot2003}, \cite{james2002}, \cite{hu2004}, \cite{muller2005}, \cite{amato2006}, \cite{hall2007}, \cite{ferraty2009}, \cite{cook2010}, \cite{chen2011}, \cite{dou2012}, \cite{febrero2013}, and \cite{goia2015}. For the case ($iii$), see \cite{ramsay1991}, \cite{fan1999}, \cite{senturk2005, senturk2008}, \cite{yao2005}, \cite{harezlak2007}, \cite{matsui2009}, \cite{valderrama2010}, \cite{he2010}, \cite{jiang2011}, \cite{ivanescu2015}, \cite{chiou2016} and \cite{zhang2018}. Also, \cite{mullerS2005}, \cite{horvath2012}, and \cite{cuevas2014} present an excellent overview of research on functional regression models and their applications. In this study, we consider the case ($iii$), which is called function-on-function regression. As a summary, the interval-valued functional regression models proposed in this study work as follows. First, the discretely observed interval-valued data are converted to functional form using a B-spline basis, and the function-on-function regression model is used to investigate the relationships between the intervals of the response and predictors. The parameter surfaces are estimated using the maximum likelihood (ML) method, and finally, the lower and upper limit functions of the response variable are obtained using the smoothing step. Throughout this study, the primary attention is paid for the prediction performance of the proposed methods. The finite sample performance of the proposed methods is evaluated numerically via Monte Carlo simulations and an empirical data example.

The remainder of this paper is organized as follows. Section~\ref{sec:methodology} reviews the classical interval-valued data regression models considered in this study and discusses their functional data extensions. In Section~\ref{sec:numerical}, Monte Carlo simulations are conducted to examine the finite sample performance of the proposed methods. An empirical data is analyzed, and the results are reported in Section~\ref{sec:real}. Section~\ref{sec:conc} concludes the paper.

\section{Methodology}\label{sec:methodology}

For $i = 1, \cdots, n$ and $j = 1, \cdots, p$, consider the following linear regression model with $p$ predictor variables $\pmb{X} = \left( \pmb{X}_1, \cdots, \pmb{X}_p \right)^\top$; $\pmb{X}_{ij} = \left(1, X_{i1}, \cdots, X_{ip} \right)^\top$ and a response variable $\pmb{Y} =\left( Y_1, \cdots, Y_n \right)^\top$, which are the realizations of intervals so that $X_{ij} = \left[ X^l_{ij}, X^u_{ij} \right]$; $X^l_{ij} \leq X^u_{ij}$ and $Y_i = \left[ Y^l_i, Y^u_i \right]$; $Y^l_i \leq Y^u_i$:
\begin{equation} 
\pmb{Y} = \pmb{X} \pmb{\beta} + \pmb{\epsilon}, \label{reg}
\end{equation}
where $\pmb{\beta} = \left( \beta_0, \beta_1, \cdots, \beta_p \right)^\top$ is a $(p + 1) \times 1$ vector of unknown parameters and $\pmb{\epsilon} = \left( \epsilon_1, \cdots, \epsilon_n \right)^\top$ denotes the error vector of dimension $n \times 1$. The first extension of the regression model given in~\eqref{reg} to the interval-valued data was suggested by \cite{billard2000}. Their extension, called center method (CM), uses the mid-points of the intervals to fit a regression model as follows:
\begin{equation}
\pmb{Y}^c = \pmb{X}^c \pmb{\beta}^c + \pmb{\epsilon}^c, \label{cm}
\end{equation}
where $Y^c_i = \left( Y^l_i + Y^u_i\right) / 2$ and $X^c_{ij} = \left( X^l_{ij} + X^u_{ij} \right) / 2$, for $i = 1, \cdots, n$ and $j = 1, \cdots, p$. The model parameters $\pmb{\beta}^c$ are estimated using least squares (LS) method. Let $\widehat{\pmb{\beta}}^c = \left(\widehat{\pmb{\beta}}^c_0, \widehat{\pmb{\beta}}^c_1, \cdots, \widehat{\pmb{\beta}}^c_p \right)^\top$ denote the estimate of $\pmb{\beta}^c$, then the lower and upper limits of the interval $\pmb{Y} = \left[ \pmb{Y}^l, \pmb{Y}^u \right]$ are predicted as follows:
\begin{equation*}
\widehat{\pmb{Y}}^l = \pmb{X}^l \widehat{\pmb{\beta}}^c, \qquad \widehat{\pmb{Y}}^u = \pmb{X}^u \widehat{\pmb{\beta}}^c.
\end{equation*}
However, the CM does not take into account the internal variations of the intervals when estimating the model parameters, see \cite{lima2008} and \cite{ahn2012}.

To overcome this problem, \cite{lima2004} suggested the center-range method (CRM) by fitting two distinct regression models using the mid-points (as in~\eqref{cm}) and half-ranges of the intervals. Let $Y^r_i = \left( Y^u_i - Y^l_i \right) / 2$ and $X^r_{ij} = \left( X^u_{ij} - X^l_{ij} \right) / 2$, for $i = 1, \cdots, n$ and $j = 1, \cdots, p$, denote the half-ranges of the intervals of $\left( X_1, \cdots, X_p \right)$ and $Y$, respectively. Then, the regression equation of the half-ranges is given as follows:
\begin{equation}
\pmb{Y}^r = \pmb{X}^r \pmb{\beta}^r + \pmb{\epsilon}^r, \label{radius}
\end{equation}
where $\pmb{\beta}^r = \left( \beta^r_0, \beta^r_1, \cdots, \beta^r_p \right)^\top$ and $ \pmb{\epsilon}^r = \left( \epsilon^r_1, \cdots, \epsilon^r_n \right)^\top$. The model parameters are estimated by the LS method. Let $\widehat{\pmb{Y}}^r = \pmb{X}^r \pmb{\widehat{\beta}}^r$ denote the predicted half-ranges of the intervals. Then, the predictions of the lower and upper limits of the response variable are obtained as follows:
\begin{equation*}
\widehat{\pmb{Y}}^l = \widehat{\pmb{Y}}^c - \widehat{\pmb{Y}}^r, \qquad \widehat{\pmb{Y}}^u = \widehat{\pmb{Y}}^c + \widehat{\pmb{Y}}^r.
\end{equation*}
As pointed out by \cite{ahn2012}, the CRM assumes that mid-points and half-ranges are independent, which may not hold in general. 

\cite{billardbook} proposed bivariate center and range method (BRCM) to take into account the effects of intervals widths. It constructs two distinct regression models using mid-points and half-ranges of the intervals. However, unlike the CRM, both the mid-points and half-ranges are used as predictors in the regression models. Let $X^{cr}_i = \left(1, X^c_{i1}, \cdots, X^c_{ip}, X^r_{i1}, \cdots, X^r_{ip} \right)^\top$. Then, the regression equations for the BCRM are given as follows:
\begin{equation}\label{bcrm}
\pmb{Y}^c = \pmb{X}^{cr} \pmb{\beta}^c + \pmb{\epsilon}^c, \qquad \pmb{Y}^r = \pmb{X}^{cr} \pmb{\beta}^r + \pmb{\epsilon}^r, 
\end{equation}
where $\pmb{\beta}^c = \left( \beta_0^c, \beta_1^c, \cdots, \beta_{2p}^c \right)^\top$ and $\pmb{\beta}^r = \left( \beta_0^r, \beta_1^r, \cdots, \beta_{2p}^r \right)^\top$ denote the parameter vectors. As in CRM, the regression parameters are estimated using the LS method, accordingly the lower and upper limits of the response variable are predicted as follows:
\begin{equation*}
\widehat{\pmb{Y}}^l = \widehat{\pmb{Y}}^c - \widehat{\pmb{Y}}^r, \qquad \widehat{\pmb{Y}}^u = \widehat{\pmb{Y}}^c + \widehat{\pmb{Y}}^r.
\end{equation*}

All three regression models (CM, CRM, and BRCM), are not appropriate for statistical inference, such as coefficient and model significance tests. \cite{ahn2012} proposed a resampling based interval-valued data regression model (MCM), which enables making inferences on the model. Let $B$ denote the number of Monte Carlo simulations. Then, the MCM works as follows:
\begin{itemize}
\item[Step 1.] Generate single-valued predictor/s $X^*_{ij}$ and response $Y^*_i$ variables uniformly from the intervals $X_{ij} = \left[ X^l_{ij}, X^u_{ij} \right]$ and $Y_i = \left[ Y^l_i, Y^u_i \right]$, respectively, to get a random vector $\left( Y^{*}_i, X^{*}_{i1}, \cdots, X^{*}_{ip} \right)^\top$, for $i = 1, \cdots, n$ and $j = 1, \cdots, p$.
\item[Step 2.] Construct a linear regression using the generated single-valued observations $\pmb{Y}^* = \left( Y^*_1, \cdots, Y^*_n \right)$ and $\pmb{X}^*_j = \left( X^*_{1j}, \cdots, X^*_{nj} \right) $ in Step 1 as follows:
\begin{equation*}
\pmb{Y}^* = \pmb{X}^* \pmb{\beta}^* + \pmb{\epsilon}^*.
\end{equation*}
\item[Step 3.] Estimate the regression coefficients $\widehat{\pmb{\beta}}^* = \left( \widehat{\beta}^*_0, \widehat{\beta}^*_1, \cdots, \widehat{\beta}^*_p \right)$ using the LS method:
\begin{equation*}
\widehat{\pmb{\beta}}^* = \left( \left( \pmb{X}^* \right)^\top \pmb{X}^* \right)^{-1} \left( \pmb{X}^* \right)^\top \pmb{Y}^*.
\end{equation*}
\item[Step 4.] Repeat Steps 1-3 $B$ times to obtain $B$ sets of estimates, $\widehat{\pmb{\beta}}^* = \left( \widehat{\pmb{\beta}}^{*1}, \cdots, \widehat{\pmb{\beta}}^{*B} \right)$.
\end{itemize}
Let $\overline{\widehat{\pmb{\beta}}}^* = B^{-1} \sum_{b=1}^B \widehat{\pmb{\beta}}^{*b}$ be the final estimate. Then, the lower and upper values of the response variable are predicted as follows:
\begin{equation*}
\widehat{\pmb{Y}}^l = \overline{\widehat{\pmb{\beta}}}^* \pmb{X}^l, \qquad \widehat{\pmb{Y}}^u = \overline{\widehat{\pmb{\beta}}}^* \pmb{X}^u.
\end{equation*}
Note that, for CM and MCM, the lower bound $\widehat{\pmb{Y}}^l$ may produce higher values than the upper bound $\widehat{\pmb{Y}}^u $. Thus, the authors suggested to use $\widehat{\pmb{Y}}^l = \min \left( \widehat{\pmb{Y}}^l, \widehat{\pmb{Y}}^u \right)$ and $\widehat{\pmb{Y}}^u = \max \left( \widehat{\pmb{Y}}^l, \widehat{\pmb{Y}}^u \right)$ to obtain logical predictions.

\subsection{Functional response model}\label{sec:func}

A functional data $\left\lbrace \X_i(t): i = 1, \cdots, N, ~ t \in \Tau \right\rbrace$ comprises random functions which are sample elements recorded at discrete times $t = \left\lbrace t_1, \cdots, t_J \right\rbrace \subset \Tau$. Let $\left( \mathcal{H}, \langle\cdot,\cdot\rangle \right)$ denote a separable Hilbert space, where $\langle\cdot,\cdot\rangle$ represents the inner product which generates the norm $\parallel \cdot \parallel$. Then, the functional random variable $\X$ is defined as $\X: \left( \Omega, \Sigma, P\right) \rightarrow \mathcal{H}$ where $\Omega$, $\Sigma$ and $P$ represent the sample space, $\sigma$-algebra and the probability measure, respectively. Throughout this research, we consider $L^2 \left( \Tau \right)$-Hilbert space of square integrable functions $f: \Tau \rightarrow \mathbb{R}$ satisfying $\int_{\Tau} f^2(t)dt < \infty$ with an inner product $\langle f,g \rangle = \int_{\Tau} f(t) g(t) dt$, $\forall f, g \in L^2$. In addition, we assume that the random variable $\X \left(t \right) \in L^2 \left(\Omega\right)$ with finite second order moment is a second order stochastic process so that $E \left[ \vert \X \vert^2 \right] = \int_{\Omega} \vert \X \vert^2 dP < \infty$.

The functional relationship between a functional response $\Y_i(t)$ and $M$ functional predictors $\X_{im}(s)$; $\left\lbrace \left(\Y_i(t), \X_{im}(s)\right); i = 1, \cdots, N, m = 1, \cdots, M, s \in \Tau_m, t \in \Tau \right\rbrace$, where $\Tau \subset \mathbb{R}$ and $\Tau_m \subset \mathbb{R}$ are the ranges of response and predictors, respectively, can be formulated by the following multiple functional linear regression model \citep{ramsay2005}:
\begin{equation}
\Y_i(t) = \beta_0(t) + \sum_{m=1}^M \int_{\Tau_m} \X_{im}(s) \beta_m(s,t) ds + \epsilon_i(t), \label{reg1}
\end{equation}
where $\beta_0(t)$ is the mean function, $\beta_m(s,t)$ is the bivariate coefficient function of the $m$\textsuperscript{th} predictor, and $\epsilon_i(t)$ is the random error function. Without loss of generality, the mean function can be eliminated from the model~\eqref{reg1} by centering the functional response and predictors as follows:
\begin{eqnarray}\label{cent}
\Y^*_i(t) &=& \Y_i(t) - \overline{\Y}(t), \nonumber \\
\X^*_{im}(s) &=& \X_{im}(s) - \overline{\X}_m(s), \nonumber
\end{eqnarray}
where $\overline{\Y}(t) = N^{-1} \sum_{i=1}^N \Y_i(t)$ and $\overline{\X}_m(s) = N^{-1} \sum_{i=1}^N \X_{im}(s)$ denote the mean of the functional response and $m$\textsuperscript{th} functional predictor, respectively. Consequently, the functional regression model~\eqref{reg1} can be re-written as follows:
\begin{equation}
\Y^*_i(t) = \sum_{m=1}^M \int_{\Tau_m} \X^*_{im}(s) \beta_m(s,t) ds + \epsilon^*_i(t), \label{regc}
\end{equation}
where $\epsilon^*_i(t) = \epsilon_i(t) - \overline{\epsilon}(t)$ is the centered error function. The estimation process of the regression coefficient function $\beta_m(s,t)$ consists of three steps: ($i$) First, discretize the estimation problem, ($ii$) solve for a matrix, say $\pmb{B}$, and ($iii$) apply smoothing step to obtain coefficient function.

A function $x(t)$ can be approximated by a linear combination of basis functions and associated coefficients for a sufficiently large number of basis functions $K$: 
\begin{equation*}
x(t) = \sum_{k=1}^K c_k \phi_k(t),
\end{equation*} 
where, for $k = 1, \cdots, K$, $\phi_k(t)$ and $c_k$ denote the basis functions and corresponding coefficients, respectively. The popular basis functions include Fourier, B-spline, and Gaussian basis functions (\cite{ramsay2005} and \cite{matsui2009}). In this study, we consider the B-spline basis to approximate functions.

The centered functional response and functional predictors, as well as the bivariate coefficient function, can be written as basis function expansions as follows:
\begin{eqnarray}\label{eq:smv}
\Y^*_i(t) &=& \sum_{k=1}^{K_\Y} c_{ik} \phi_k(t) = \pmb{c}^\top_i \pmb{\Phi}(t), \qquad \forall t \in \Tau, \nonumber \\
\X^*_{im}(s) &=& \sum_{j=1}^{K_{m,\X}} d_{imj} \psi_{mj}(s) = \pmb{d}^\top_{im} \pmb{\Psi}_m(s), \qquad \forall s \in \Tau_m, \nonumber \\
\beta_m(s,t) &=& \sum_{j,k} \psi_{mj}(s) b_{mjk} \phi_k(t) = \pmb{\Psi}^\top_m(s) \pmb{B}_m \pmb{\Phi}(t), \nonumber
\end{eqnarray}
where $\pmb{\Phi}(t) = \left( \phi_1(t), \cdots, \phi_{K_\Y}(t) \right)^\top$ and $\pmb{\Psi}_m(s) = \left( \psi_{m1}(s), \cdots, \psi_{mK_{m,\X}}(s) \right)^\top$ represent the vector of basis functions, $\pmb{c}_{i} = \left( c_{i1}, \cdots, c_{iK_\Y} \right)^\top$ and $\pmb{d}_{im} = \left( d_{im1}, \cdots, d_{imK_{m,\X}} \right)^\top$ are the corresponding coefficient vectors, and $\pmb{B}_m = (b_{mjk})_{j,k}$ denotes the coefficient matrix of dimension $K_{m,\X} \times K_\Y$. Accordingly, the regression model~\eqref{regc} can be re-expressed as follows:
\begin{eqnarray}\label{regs}
\pmb{c}^\top_i \pmb{\Phi}(t) &=& \sum_{m=1}^M \pmb{d}^\top_{im} \pmb{\zeta}_{\psi_m} \pmb{B}_m \pmb{\Phi}(t) + \epsilon^*_i(t), \nonumber \\
&=& \pmb{z}^\top_i \pmb{B} \pmb{\Phi}(t) + \epsilon^*_i(t),
\end{eqnarray}
where $\pmb{\zeta}_{\psi_m} = \int_{\Tau_m} \pmb{\Psi}_m(s) \pmb{\Psi}^\top_m(s) ds$ is a $K_{m,\X} \times K_{m,\X}$ matrix, $\pmb{z}_i = \left( \pmb{d}^\top_{i1} \pmb{\zeta}_{\psi_1}, \cdots, \pmb{d}^\top_{iM} \pmb{\zeta}_{\psi_M} \right)^\top$ is a vector of length $\sum_{m=1}^M K_{m,\X}$ and $\pmb{B} = \left( \pmb{B}_1, \cdots, \pmb{B}_M \right)^\top$ is the coefficient matrix with dimension $\sum_{m=1}^M K_{m,\X} \times K_\Y$.

Suppose now that the error function $\epsilon^*_i(t)$ in \eqref{regs} can be approximated as basis function expansion, $\epsilon^*_i(t) = \pmb{e}^\top_i \pmb{\Phi}(t)$ where $\pmb{e}_i = \left( e_{i1}, \cdots, e_{iK} \right)^\top$ is a vector of independently and identically distributed Gaussian random variables with mean $\pmb{0}$ and variance-covariance matrix $\pmb{\Sigma}$. Then, the regression model in~\eqref{regs} has the following form:
\begin{equation}
\pmb{c}^\top_i \pmb{\Phi}(t) = \pmb{z}^\top_i \pmb{B} \pmb{\Phi}(t) + \pmb{e}^\top_i \pmb{\Phi}(t). \label{rmax}
\end{equation}
Multiplying both sides of equation~\eqref{rmax} from the right by $\pmb{\Phi}^\top(t)$ and integrating with respect to $\Tau$ yields:

\begin{eqnarray}
\pmb{c}^\top_i \pmb{\Phi}(t) \pmb{\Phi}^\top(t) &=& \pmb{z}^\top_i \pmb{B} \pmb{\Phi}(t) \pmb{\Phi}^\top(t) + \pmb{e}^\top_i \pmb{\Phi}(t) \pmb{\Phi}^\top(t), \nonumber \\
\int_{\Tau} \pmb{c}^\top_i \pmb{\Phi}(t) \pmb{\Phi}^\top(t) dt &=& \int_{\Tau} \pmb{z}^\top_i \pmb{B} \pmb{\Phi}(t) \pmb{\Phi}^\top(t) dt + \int_{\Tau} \pmb{e}^\top_i \pmb{\Phi}(t) \pmb{\Phi}^\top(t) dt, \nonumber \\
\pmb{c}^\top_i \pmb{\zeta}_{\Phi} &=& \pmb{z}^\top_i \pmb{B} \pmb{\zeta}_{\Phi} + \pmb{e}^\top_i \pmb{\zeta}_{\Phi}, \nonumber \\
\pmb{c}_i &=& \pmb{B}^\top \pmb{z}_i + \pmb{e}_i, \nonumber
\end{eqnarray}
since $\pmb{\zeta}_{\Phi}$ is non-singular. The probability density function of a functional response $\Y_i$ given a functional predictor $\X_i$ and parameter vector $\pmb{\theta} = \left( \pmb{B}, \pmb{\Sigma} \right)$ can be written as follow:
\begin{equation*}
f \left( \Y_i | \X_i; \pmb{\theta} \right) = \frac{1}{\left( 2 \pi \right)^{K/2} \vert \pmb{\Sigma} \vert^{1/2}} \exp \left\lbrace  - \frac{1}{2} \left( \pmb{c}_i - \pmb{B}^\top \pmb{z}_i \right)^\top \pmb{\Sigma}^{-1} \left( \pmb{c}_i - \pmb{B}^\top \pmb{z}_i \right) \right\rbrace,
\end{equation*}
since $\pmb{e}_i \sim \pmb{N} \left( \pmb{0}, \pmb{\Sigma} \right)$. Then, the $\log$-likelihood function $\ell \left( \Y_i | \X_i; \pmb{\theta} \right) = \sum_{i=1}^N \log f \left( \Y_i | \X_i; \pmb{\theta} \right)$ is obtained as:
\begin{equation*} \label{loglik}
\ell \left( \Y_i | \X_i; \pmb{\theta} \right) = - \frac{N}{2} \log \vert \pmb{\Sigma} \vert - \frac{1}{2} tr \left\lbrace \pmb{\Sigma}^{-1} \left( \pmb{C} - \pmb{Z} \pmb{B} \right)^\top \left( \pmb{C} - \pmb{Z} \pmb{B} \right) \right\rbrace,
\end{equation*}
where $\pmb{C} = \left( \pmb{c}_1, \cdots, \pmb{c}_N \right)^\top$ and $\pmb{Z} = \left( \pmb{z}_1, \cdots, \pmb{z}_N \right)^\top$. By equating the derivatives of the $\log$-likelihood function with respect to $\pmb{\theta} = \left( \pmb{B}, \pmb{\Sigma} \right)$ to $\pmb{0}$, the ML estimators $\widehat{\pmb{\theta}} = \left( \widehat{\pmb{B}}, \widehat{\pmb{\Sigma}} \right)$ are obtained as follows:
\begin{eqnarray*}
\widehat{\pmb{B}} &=& \left( \pmb{Z}^\top \pmb{Z} \right)^{-1} \pmb{Z} \pmb{C}, \\
\widehat{\pmb{\Sigma}} &=& \frac{1}{N} \left( \pmb{C} - \pmb{Z} \widehat{\pmb{B}} \right)^\top \left( \pmb{C} - \pmb{Z} \widehat{\pmb{B}} \right).
\end{eqnarray*} 
Finally, the fitted values of the response function $\Y(t) = \left( \Y_1(t), \cdots, \Y_N(t) \right)^\top$ are obtained as follows:
\begin{equation*}
\widehat{\Y}(t) = \left( \pmb{Z} \widehat{\pmb{B}} \right) \pmb{\Phi}(t) + \overline{\Y}(t).
\end{equation*}

\subsection{Functional response model for interval-valued functional data}

The interval-valued functional data consist of functions that define the interval-valued data as functions. In other words, an interval-valued functional data consist of two functions; a lower limit function $\X^l(t)$ and an upper limit function $\X^u(t)$ such that $\X(t) = \left( \X^l(t), \X^u(t) \right)$, $\X^l(t) \leq \X^u(t)$. Denote by $\X^c(t) = \left( \X^u(t) - X^l(t) \right) / 2$ and $\X^r(t) = \left( \X^u(t) - X^l(t) \right) / 2$ the center and half-range functions, respectively. Let $\pmb{\Phi}^l(t) = \left( \phi^l_1(t), \cdots, \phi^l_{K^l}(t) \right)$, $\pmb{\Phi}^u(t) = \left( \phi^u_1(t), \cdots, \phi^u_{K^u}(t) \right)$, $\pmb{\Phi}^c(t) = \left( \phi^c_1(t), \cdots, \phi^c_{K^c}(t) \right)$, and $\pmb{\Phi}^r(t) = \left( \phi^r_1(t), \cdots, \phi^r_{K^r}(t) \right)$ denote the vectors of basis functions of the lower limit, upper limit, center, and half-range functions, respectively. Then, they can be expressed as the linear combinations of basis functions as follows:
\begin{equation*}
\X^l(t) = \sum_{k=1}^{K^l} c^l_{k} \phi^l_{k}(t), \qquad X^u(t) = \sum_{k=1}^{K^u} c^u_{k} \phi^u_{k}(t), \qquad X^r(t) = \sum_{k=1}^{K^r} c^r_{k} \phi^r_{k}(t), \qquad \qquad X^c(t) = \sum_{k=1}^{K^c} c^c_{k} \phi^c_{k}(t),
\end{equation*}
where $c^l_{k}$, $c^u_{k}$, $c^c_{k}$, and $c^r_{k}$ denote the coefficient vectors of the lower limit, upper limit, center, and half-range functions, respectively. 

Before introducing the functional forms of the interval-valued data regression models, we note that for an interval-valued functional data $\X(t) = \left( \X^l(t), \X^u(t) \right)$, one needs to find a common vector of basis functions $\pmb{\Phi}^{cm}(t)$ that works for all the components in the interval; lower limit, upper limit, center, and half-range functions. Otherwise, the regression model may not be constructed since different components may have a different number of basis functions. 

Suppose now that the (centered) interval-valued random functional response $\Y^*_i(t)$ and functional predictors $\X^*_{im}(s)$, for $i = 1, \cdots, N$ and $m = 1, \cdots, M$, be the realizations of the intervals $\Y^*_i(t) = \left( \Y^{*l}_i(t), \Y^{*u}_i(t) \right)$ and $\X^*_{im}(s) = \left( \X^{*l}_{im}(s), \X^{*u}_{im}(s) \right)$. Denote by $\Y^{*c}_i(t)$ and $\X^{*c}_{im}(s)$ the center functions of the response and $m$\textsuperscript{th} predictor variables, respectively.  Then, the CM defined in~\eqref{cm} can be extended to the functional data as follows:
\begin{equation}
\Y^{*c}_i(t) = \sum_{m=1}^M \int_{\Tau_m} \X^{*c}_{im}(s) \beta^c_m(s,t) ds + \epsilon^{*c}_i(t). \label{fcm}
\end{equation}
Let $\pmb{C}^c = \left(\pmb{B}^c\right)^\top \pmb{Z}^c + \pmb{e}^c$ denote the multivariate regression model constructed using the basis function expansions of the functional objects given in~\eqref{fcm} and following Section~\eqref{sec:func}. Let also $\widehat{\pmb{B}}^c$ denote the ML estimate of  $\pmb{B}^c$. Then, the predictions of the lower and upper limit functions of the interval-valued functional response variable using the functional CM are obtained as follows: 
\begin{equation*}
\widehat{\Y}^l(t) = \left( \pmb{Z}^l \widehat{\pmb{B}}^c \right) \pmb{\Phi}^{cm}(t) + \overline{\Y}^l(t), \qquad \widehat{\Y}^u(t) = \left( \pmb{Z}^u \widehat{\pmb{B}}^c \right) \pmb{\Phi}^{cm}(t) + \overline{\Y}^u(t),
\end{equation*}
where $\overline{\Y}^l(t) = N^{-1} \sum_{i=1}^N \Y^l_i(t)$, $\overline{\Y}^u(t) = N^{-1} \sum_{i=1}^N \Y^u_i(t)$, $\pmb{Z}^l$ and $\pmb{Z}^u$ denote the matrices (see~\eqref{regs}) obtained using the basis function expansions of the lower and upper limit functions of the predictors, respectively, and $\pmb{\Phi}^{cm}(t)$ is the vector of basis functions used to approximate the components of the response variable.

Similarly to the traditional case, the functional CRM is based on two distinct function-on-function regression models: the regression models of the center and half-range functions of the variables. Let $\Y^{*r}_i(t) = \left( \Y^{*u}_i(t) - \Y^{*l}_i(t)\right)  / 2$ and $\X^{*r}_{im}(s) = \left( \X^{*u}_{im}(s) - \X^{*l}_{im}(s) \right) / 2$ denote the half-range functions of the centered functional response and predictor variables, respectively. Then, the functional regression equation of the functional half-range variables is given by:
\begin{equation*}
\Y^{*r}_i(t) = \sum_{m=1}^M \int_{\Tau_m} \X^{*r}_{im}(s) \beta^r_m(s,t) ds + \epsilon^{*r}_i(t). \label{fhr}
\end{equation*}
Denote by $\pmb{C}^r = \left(\pmb{B}^r\right)^\top \pmb{Z}^r + \pmb{e}^r$ the multivariate regression model obtained using the basis function expansions of the half-range functions, and let $\widehat{\pmb{B}}^r$ be the ML estimate of $\pmb{B}^r$. Then, the predicted half-range of the functional response variable is obtained as follows:
\begin{equation*}
\widehat{\Y}^r(t) = \left( \pmb{Z}^r \widehat{\pmb{B}}^r \right) \pmb{\Phi}^{cm}(t) + \overline{\Y}^r(t),
\end{equation*}
where $\pmb{Z}^r$ is a matrix obtained by the basis function expansions of the half-range functions of the predictor variables and $\overline{\Y}^r(t) = N^{-1} \sum_{i=1}^N \Y^r_i(t)$. Subsequently, the predictions of the lower and upper limit functions of the response variable using the functional CRM are obtained as follows:
\begin{equation*}
\widehat{\Y}^l(t) = \widehat{\Y}^c(t) - \widehat{\Y}^r(t), \qquad \widehat{\Y}^u(t) = \widehat{\Y}^c(t) + \widehat{\Y}^r(t),
\end{equation*}
where $\widehat{\Y}^c(t) = \left( \pmb{Z}^c \widehat{\pmb{B}}^c \right) \pmb{\Phi}^{cm}(t) + \overline{\Y}^c(t)$ with $\overline{\Y}^c(t) = N^{-1} \sum_{i=1}^N \Y^c_i(t)$.

The functional extension of the BCRM can be obtained similarly to~\eqref{bcrm} but using the functional variables as follows:
\begin{eqnarray}
\Y^{*c}_i(t) &=& \sum_{m=1}^M \int_{\Tau_m} \X^{*cr}_{im}(s) \beta^{ccr}_m(s,t) ds + \epsilon^{*ccr}_i(t), \label{fbcrm1} \\
\Y^{*r}_i(t) &=& \sum_{m=1}^M \int_{\Tau_m} \X^{*cr}_{im}(s) \beta^{rcr}_m(s,t) ds + \epsilon^{*rcr}_i(t), \label{fbcrm2}
\end{eqnarray}
where $\X^{*cr}_{im}(s) = \left( \X^{*c}_{im}(s), \X^{*r}_{im}(s) \right)$, for $i = 1, \cdots, N$ and $m = 1, \cdots, M$. Let us denote by $\pmb{C}^{ccr} = \left(\pmb{B}^{ccr}\right)^\top \pmb{Z}^{ccr} + \pmb{e}^{ccr}$ and $\pmb{C}^{rcr} = \left(\pmb{B}^{rcr}\right)^\top \pmb{Z}^{rcr} + \pmb{e}^{rcr}$ the regression models constructed by the basis function expansions of the functional objects given in~\eqref{fbcrm1} and~\eqref{fbcrm2}, respectively. Let $\widehat{\pmb{B}}^{ccr}$ and $\widehat{\pmb{B}}^{rcr}$ denote the ML estimates of $\pmb{B}^{ccr}$ and $\pmb{B}^{rcr}$, respectively. Then, the predictions of the lower and upper limit functions of the response variable using the functional BCRM are obtained as follows:
\begin{equation*}
\widehat{\Y}^l(t) = \widehat{\Y}^c(t) - \widehat{\Y}^r(t), \qquad \widehat{\Y}^u(t) = \widehat{\Y}^c(t) + \widehat{\Y}^r(t),
\end{equation*}
where $\widehat{\Y}^c(t) = \left( \pmb{Z}^{crc} \widehat{\pmb{B}}^{crc} \right) \pmb{\Phi}^{cm}(t) + \overline{\Y}^c(t)$ and $\widehat{\Y}^r(t) = \left( \pmb{Z}^{rrc} \widehat{\pmb{B}}^{rrc} \right) \pmb{\Phi}^{cm}(t) + \overline{\Y}^c(t)$.

To extend the traditional MCM to the functional case, we provide the following algorithm.
\begin{itemize}
\item[Step 1.] For $i = 1, \cdots, N$ and $m = 1, \cdots, M$, generate a functional response $\Y_i(t)$ and $N$ sets of $M$ functional predictors $\X_{im}(s)$ uniformly from their intervals, $\left( \Y^{l}_i(t), \Y^{u}_i(t) \right)$ and $\left( \X^{l}_{im}(s), \X^{u}_{im}(s) \right)$.
\item[Step 2.] Center the generated functional response and functional predictors and construct a functional regression model as follows:
\begin{equation*}
\Y^*_i(t) = \sum_{m=1}^M \int_{\Tau_m} \X^*_{im}(s) \beta_m(s,t) ds + \epsilon^*_i(t).
\end{equation*}
\item[Step 3.] Obtain the multivariate regression model $\pmb{C} = \left(\pmb{B}\right)^\top \pmb{Z} + \pmb{e}$ using the basis function expansions of the centered functional objects. Then, apply the ML method to estimate the regression coefficients matrix, $\widehat{\pmb{B}} = \left( \pmb{Z}^\top \pmb{Z} \right)^{-1} \pmb{Z} \pmb{C}$ as explained in Section~\ref{sec:func}.
\item[Step 4.] Repeat Steps 1-3 $B$ times to obtain $B$ sets of coefficient matrices $\widehat{\pmb{B}}^{\text{MCM}} = \left( \widehat{\pmb{B}}^1, \cdots, \widehat{\pmb{B}}^B \right)$.
\end{itemize}
Let $\overline{\widehat{\pmb{B}}}^{\text{MCM}} = B^{-1} \sum_{b=1}^B \widehat{\pmb{B}}^b$ be the final estimate of the functional MCM. Then, the lower and upper limit functions of the response variable are predicted as follows:
\begin{equation*}
\widehat{\Y}^l(t) = \left( \pmb{Z}^l \overline{\widehat{\pmb{B}}}^{\text{MCM}} \right) \pmb{\Phi}^{cm}(t) + \overline{\Y}^l(t), \qquad \widehat{\Y}^u(t) = \left( \pmb{Z}^u \overline{\widehat{\pmb{B}}}^{\text{MCM}} \right) \pmb{\Phi}^{cm}(t) + \overline{\Y}^u(t).
\end{equation*}

Similarly to the traditional case, for the functional CM and MCM, the calculated lower limit functions of the response variable, $\widehat{\Y}^l_i(t)$ may be greater than the upper limit functions, $\widehat{\Y}^u_i$. To overcome this problem, we recommend to take the pointwise minimum and maximum values of the functions as follows:
\begin{equation*}
\widehat{\Y}^l_i(t_j) = \min \left( \widehat{\Y}^l_i(t_j), \widehat{\Y}^u_i(t_j) \right), \qquad \widehat{\Y}^u_i(t_j) = \max \left( \widehat{\Y}^l_i(t_j), \widehat{\Y}^u_i(t_j) \right),
\end{equation*}
for $i = 1, \cdots, N$ and $j = 1, \cdots, J$.

The functional MCM can also be used to construct prediction intervals for the lower and upper limit functions of the response variable. Let $\widehat{\epsilon}^l(t) = \Y^l(t) - \widehat{\Y}^l(t)$ and $\widehat{\epsilon}^u = \Y^u(t) - \widehat{\Y}^u(t)$, respectively, denote the estimated error functions for the lower and upper limits obtained using the functional MCM. Then, for $b = 1, \cdots, B$, one can calculate $B$ sets of fitted lower $\left( \widehat{\Y}^{l,1}, \cdots, \widehat{\Y}^{l,B} \right)$ and upper $\left( \widehat{\Y}^{u,1}, \cdots, \widehat{\Y}^{u,B} \right)$ limit functions as follows:
\begin{equation*}
\widehat{\Y}^{l,b} = \left( \pmb{Z}^l \widehat{\pmb{B}}^{b} \right) \pmb{\Phi}^{cm}(t) + \epsilon^{l,*} + \overline{\Y}^l(t), \qquad \widehat{\Y}^{u,b} = \left( \pmb{Z}^u \widehat{\pmb{B}}^{b} \right) \pmb{\Phi}^{cm}(t) + \epsilon^{u,*} + \overline{\Y}^u(t),
\end{equation*}
where, $\epsilon^{l,*}$ and $\epsilon^{u,*}$ denote random samples from $\widehat{\epsilon}^l$ and $\widehat{\epsilon}^u$, respectively. Denote by $\mathbb{Q}^l_{\alpha}(t)$ and $\mathbb{Q}^u_{\alpha}(t)$ the $\alpha^{th}$ quantiles of the generated $B$ sets of MCM replicates of the fitted lower and upper limit functions, respectively. Then, the $100(1-\alpha)\%$ functional MCM prediction intervals for $\Y^l(t)$ and $\Y^u(t)$ can be computed  as $\left[ \mathbb{Q}^l_{\alpha/2}(t), \mathbb{Q}^l_{1 - \alpha/2}(t) \right]$ and $\left[ \mathbb{Q}^u_{\alpha/2}(t), \mathbb{Q}^u_{1 - \alpha/2}(t) \right]$, respectively.

\section{Numerical results}

Various Monte Carlo simulations and empirical data analysis were conducted to investigate the finite sample performance of the proposed interval-valued functional data regression models. We note that in our calculations, the generalized inverse was used to estimate the model parameter matrix $\pmb{B}$ to avoid the singular-matrix problem. All the numerical analyses were performed using \texttt{R} 3.6.0 (an example \texttt{R} code can be found at \url{https://github.com/UfukBeyaztas/FLM_interval_valued_data}).

\subsection{Simulation studies}\label{sec:numerical}

\begin{figure}[!htbp]
  \centering
  \includegraphics[width=8.5cm]{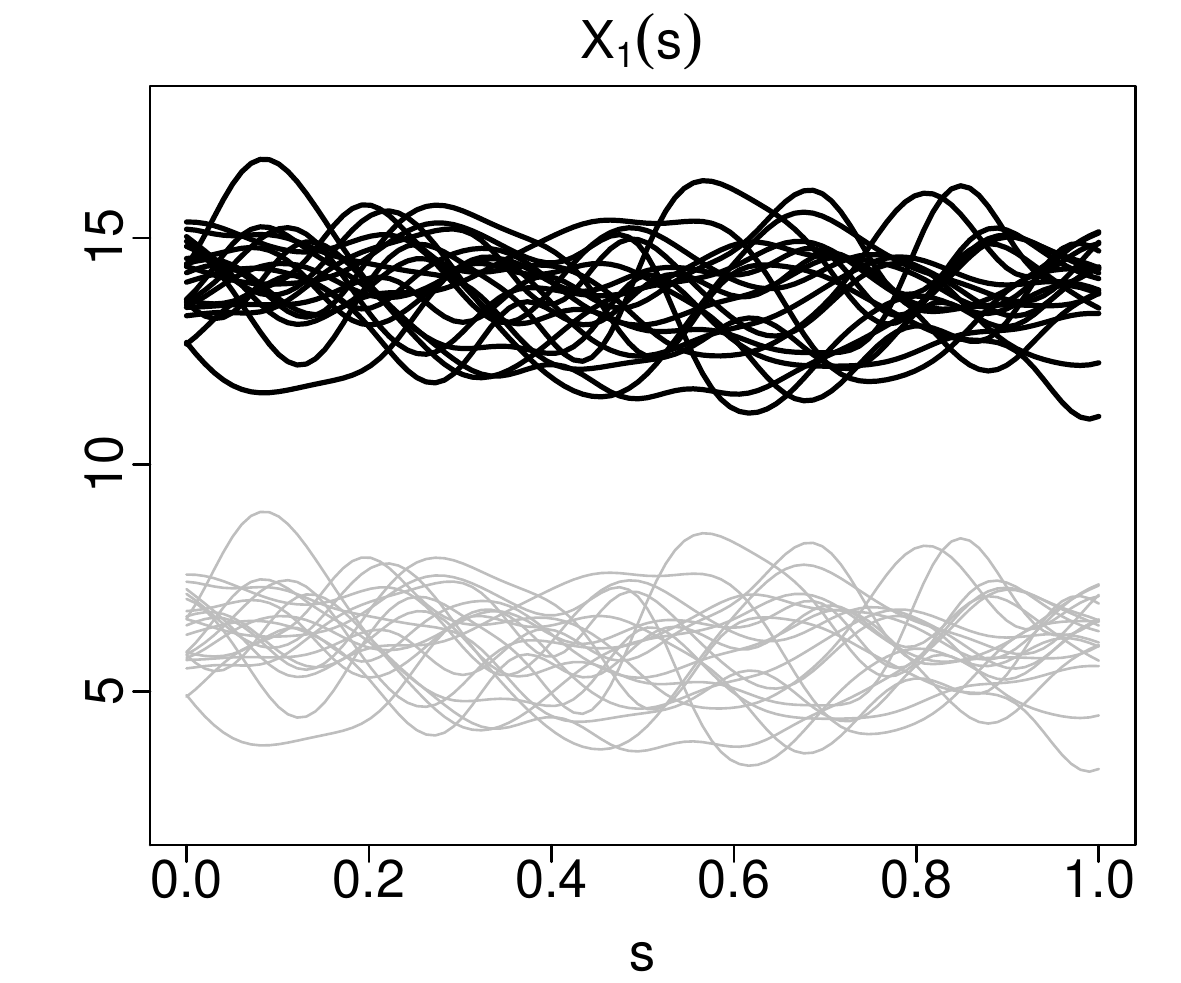}
\qquad
  \includegraphics[width=8.5cm]{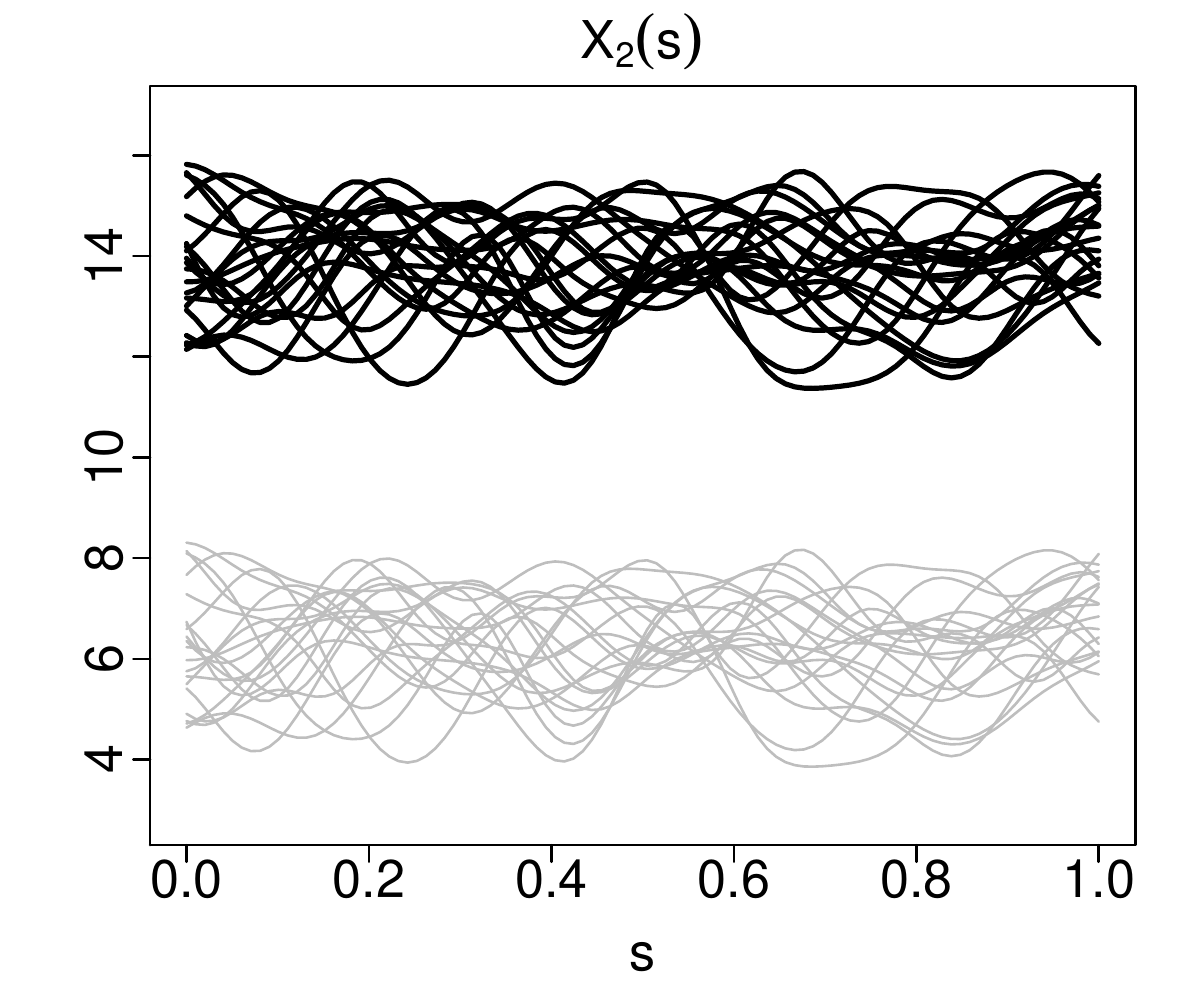}
\\  
  \includegraphics[width=8.5cm]{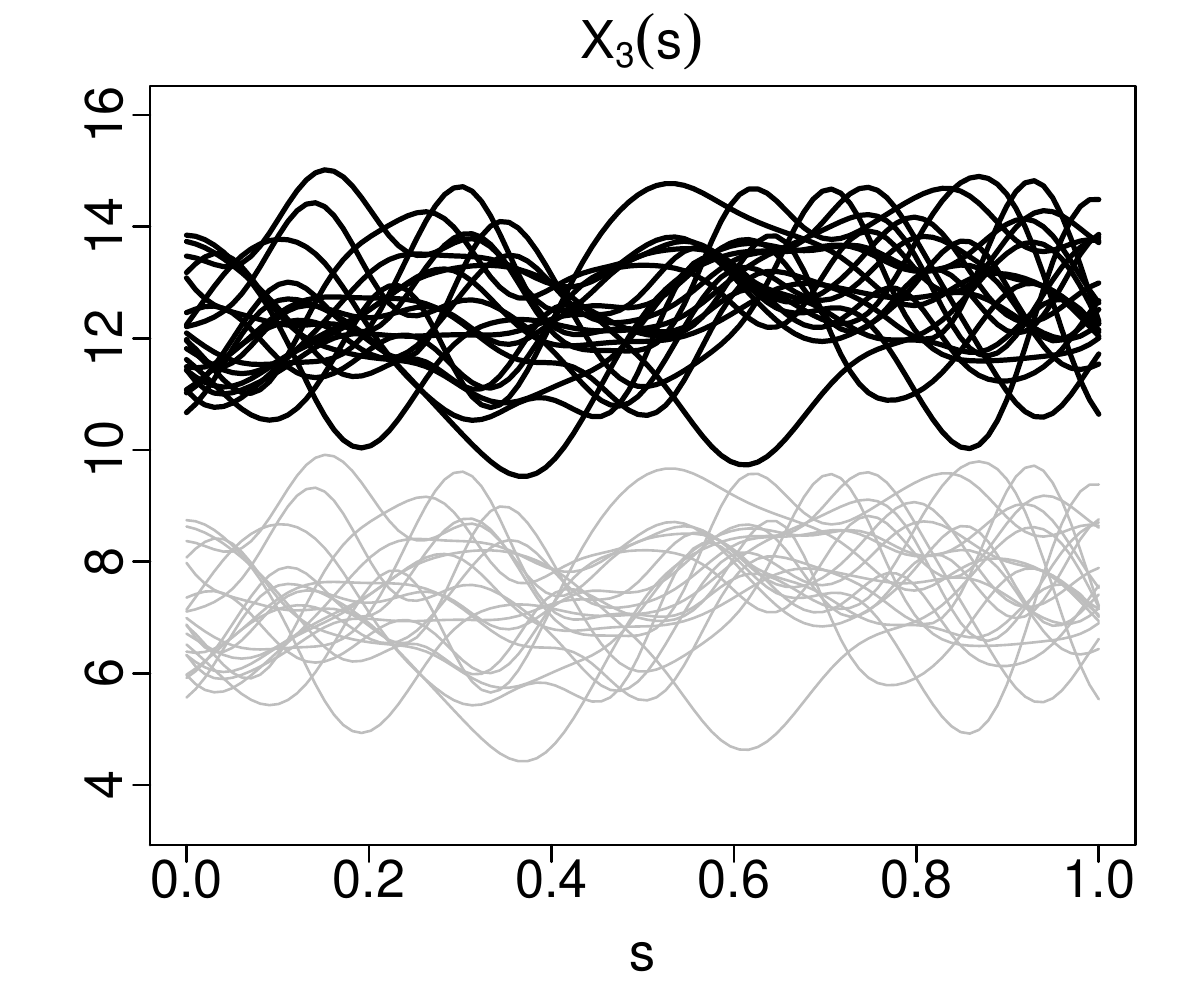}
\qquad
  \includegraphics[width=8.5cm]{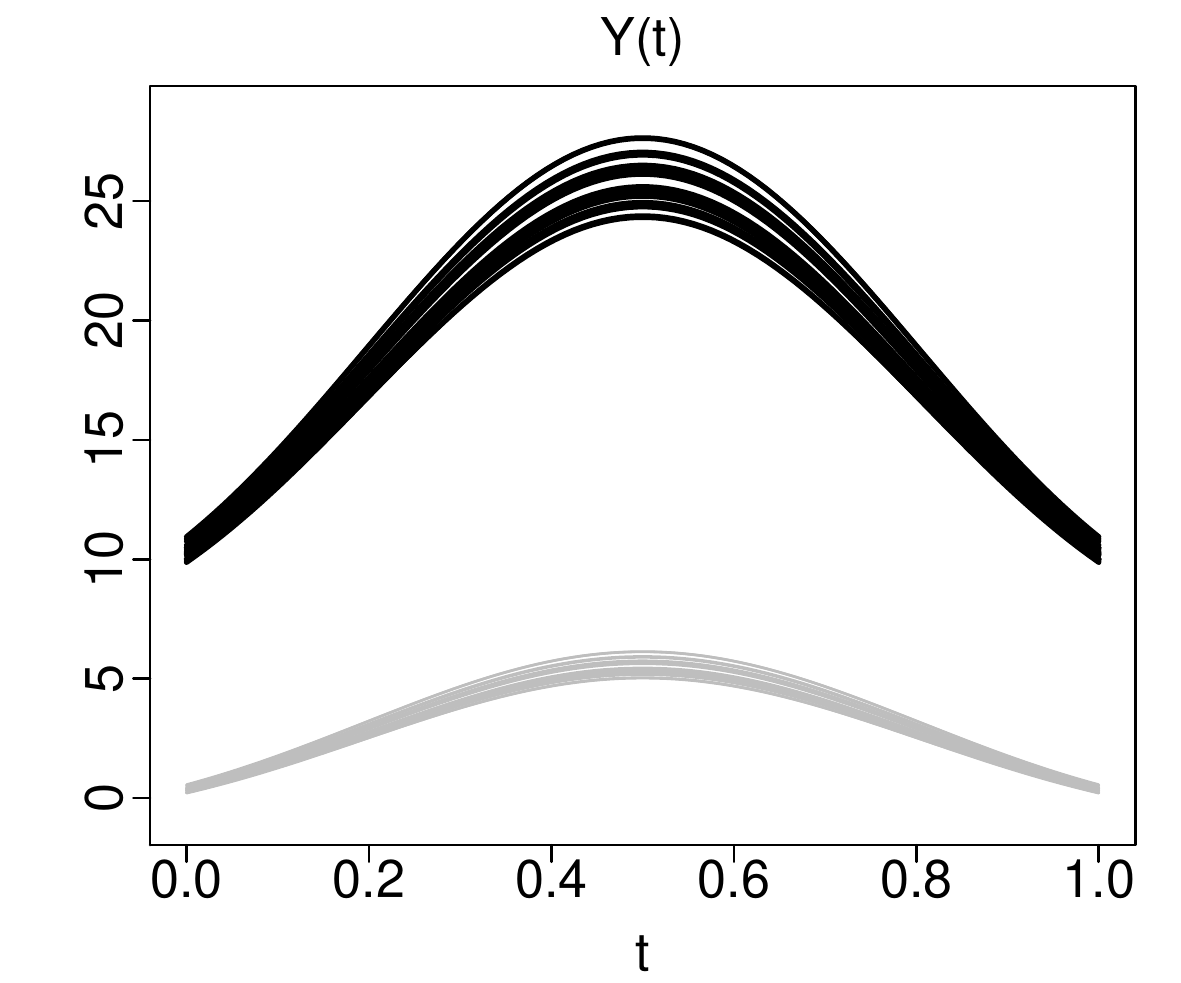}
  \caption{Plots of the generated 20 lower (gray lines) and upper (black lines) limit functions when $(a,b) = (3, 5)$ and $(c,d) = (5,8)$.}\label{fig:sim-data}
\end{figure}

Throughout the simulations, the multiple function-on-function regression model with $M = 3$ functional predictors was considered, and the following process was used to generate interval-valued functional variables.
\begin{itemize}
\item[(a)] For $i = 1, \cdots, N = 200$ and $m = 1, 2, 3$, use the following process to generate center functions of the predictors $\X_m^c(s)$; $\X_m^c(s) = 10 + V_m(s)$, where $s \in [0,1]$ and $V_m(s)$s are generated from the Gaussian process with zero mean and a positive variance-covariance function $\pmb{\Sigma}_V(s, s^{\prime}) = \exp(-100 (s - s^{\prime}))^2$.
\item[(b)] Generate the center functions of the error process; $\epsilon^c(t)$, where $t \in [0,1]$, from the normal distribution with zero mean and variance four; $N(0,4)$. Then, generate the center functions of the response variable as follows:
\begin{equation*}
\Y^c(t) = \sum_{m=1}^3 \int_0^1 \X^c_m(s) \beta^c_m(s,t) + \epsilon^c(t),
\end{equation*}
where
\begin{align*}
\beta^c_1(s,t) &= (1-s)^2 (t - 0.5)^2 \\
\beta^c_2(s,t) &= \exp(-3 (s-1)^2) \exp(-5 (t-0.5)^2) \\
\beta^c_3(sit) &= \exp(-5(s-0.5)^2 - 5(t-0.5)^2) + 8 \exp(-5(s-1.5)^2 - 5(t-0.5)^2)
\end{align*}
\item[(c)] Generate the range functions of the response $\Y^r(t)$ and predictors $\X^r_m(s)$ as follows:
\begin{equation*}
\Y^r(t) = \Y^c(t) + U(a,b), \qquad \X^r_m(s) = \X^r_m(s) + U(c,d).
\end{equation*}
For the range functions, four different cases were considered: $(a, b) = \left[ \left( 1, 1.5\right), \left(1, 3\right), \left( 3, 5 \right), \left( 8, 20 \right) \right] $ and $(c, d) = \left[ \left( 1, 1.5\right), \left( 1, 3\right), \left( 5, 8 \right), \left( 6, 15 \right) \right] $.
\item[(d)] Calculate the lower and upper limit functions of the response and predictors as follow:
\begin{align*}
\Y^l(t) &= \Y^c(t) - \Y^r(t)/2, \qquad \Y^u(t) = \Y^c(t) + \Y^r(t)/2, \\
\X^l_m(s) &= \X^c_m(s) - \X^r_m(s)/2, \qquad \X^l_m(s) = \X^c_m(s) - \X^r_m(s)/2.
\end{align*}
\end{itemize}
All the components of the functional variables were generated at 100 equally spaced points in the interval [0,1], and the components of the generated predictor variables were distorted by the Gaussian noise $u(s) \sim N(0,4)$ before fitting the interval-valued functional regression models. A graphical display of the generated lower and upper limit functions are presented in Figure~\ref{fig:sim-data}.

For each simulation setting, $\text{MC} = 250$ Monte Carlo simulations were performed. For each simulation, the first 100 functions of the generated functional variables were used to construct the interval-valued functional regression models. Note that for each simulation setting, eight number of basis functions was used to convert all the components of the generated noisy data to their functional forms. The last 100 functions of the data were used to evaluate the prediction performances of the proposed methods. The prediction performance of the methods was evaluated using the average mean squared errors (AMSE) of the lower ($\text{AMSE}^l$) and upper ($\text{AMSE}^u$) limit functions as follows:
\begin{eqnarray*}
\text{AMSE}^l &=& \frac{1}{100} \sum_{i=1}^{100} \left\Vert \Y^l_i(t) - \widehat{\Y}^l_i(t) \right\Vert_{L^2}, \\
\text{AMSE}^u &=&\frac{1}{100} \sum_{i=1}^{100} \left\Vert \Y^u_i(t) - \widehat{\Y}^u_i(t) \right\Vert_{L^2},
\end{eqnarray*}  
where $\left\Vert \cdot \right\Vert_{L^2}$ is the $L^2$ norm, which is approximated by the Riemann sum \citep{LuoQi}. Note that, for the functional MCM, the results were obtained based on a $B = 100$ re-sampling procedure. The nominal significance level $\alpha$ was set to 0.05 to construct functional MCM-based prediction intervals for the lower and upper limit functions. The coverage probabilities (CP) for the prediction intervals of lower ($\text{CP}^l$) and upper ($\text{CP}^u$) limit functions were calculated to evaluate the accuracy of the constructed prediction intervals:
\begin{align*}
\text{CP}^l &= \frac{1}{100 \times 100} \sum_{i=1}^{100} \sum_{j=1}^{100} \mathbbm{1} \left\lbrace \mathbb{Q}^{l}_{\alpha/2,i,j} \leq \Y^l_i(t_j) \leq \mathbb{Q}^{l}_{1 - \alpha/2,i,j} \right\rbrace, \\
\text{CP}^u &= \frac{1}{100 \times 100} \sum_{i=1}^{100} \sum_{j=1}^{100} \mathbbm{1} \left\lbrace \mathbb{Q}^{u}_{\alpha/2,i,j} \leq \Y^u_i(t_j) \leq \mathbb{Q}^{u}_{1 - \alpha/2,i,j} \right\rbrace,
\end{align*}
where $\mathbbm{1}\{\cdot\}$ denotes the indicator function. The finite sample performance of the proposed methods was compared with the classical functional linear model (FLM). While doing so, two functional linear models for the lower and upper limit functions of the response and predictor variables were constructed as follows:
\begin{align*}
\Y^l_i(t) &= \int_0^1 \X^l_{im}(s) \beta^l_m(s,t) ds + \epsilon^l_i(t), \\
\Y^u_i(t) &= \int_0^1 \X^u_{im}(s) \beta^u_m(s,t) ds + \epsilon^u_i(t).
\end{align*}

\begin{figure}[!htbp]
  \centering
  \includegraphics[width=8.5cm]{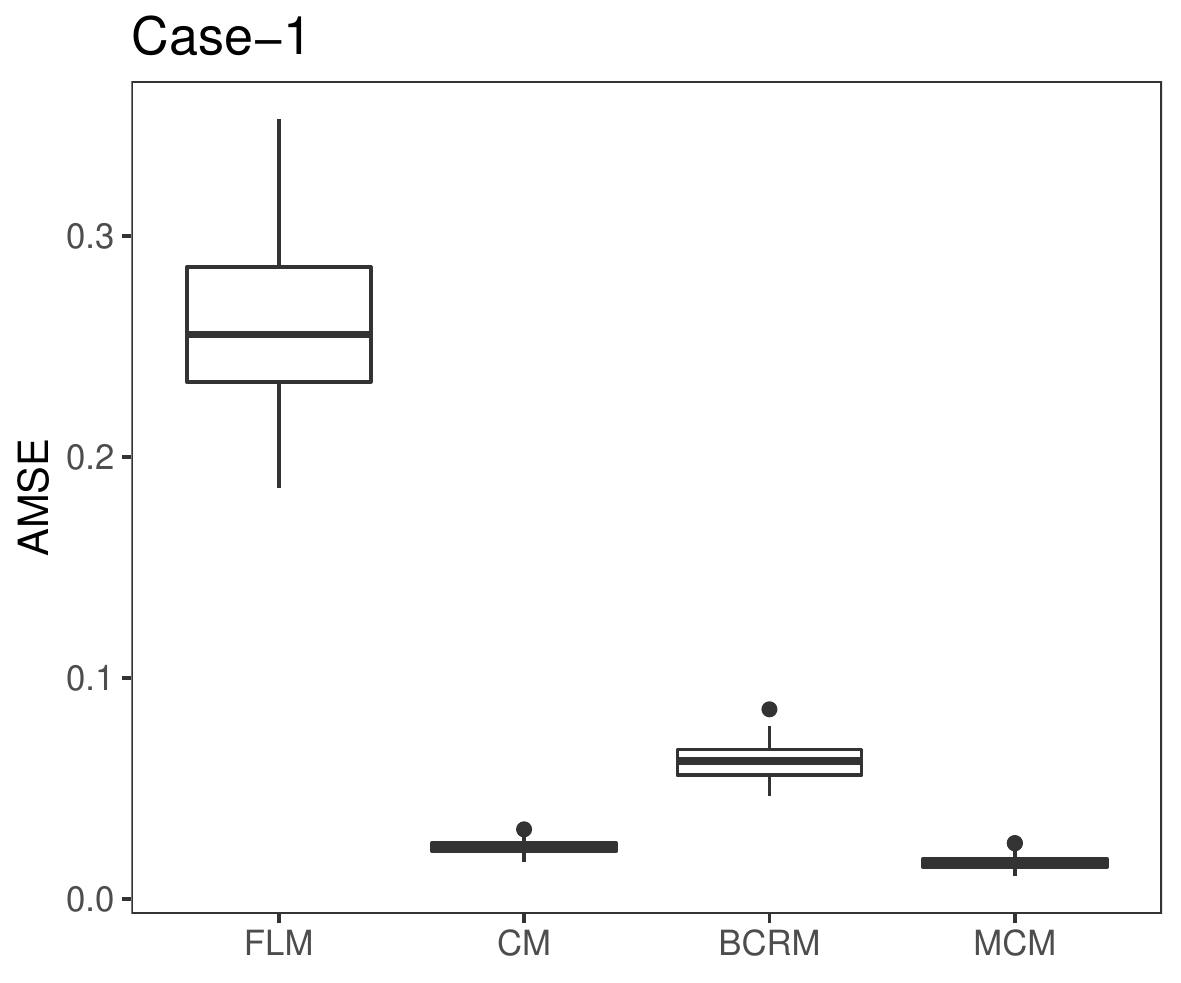}
\qquad
  \includegraphics[width=8.5cm]{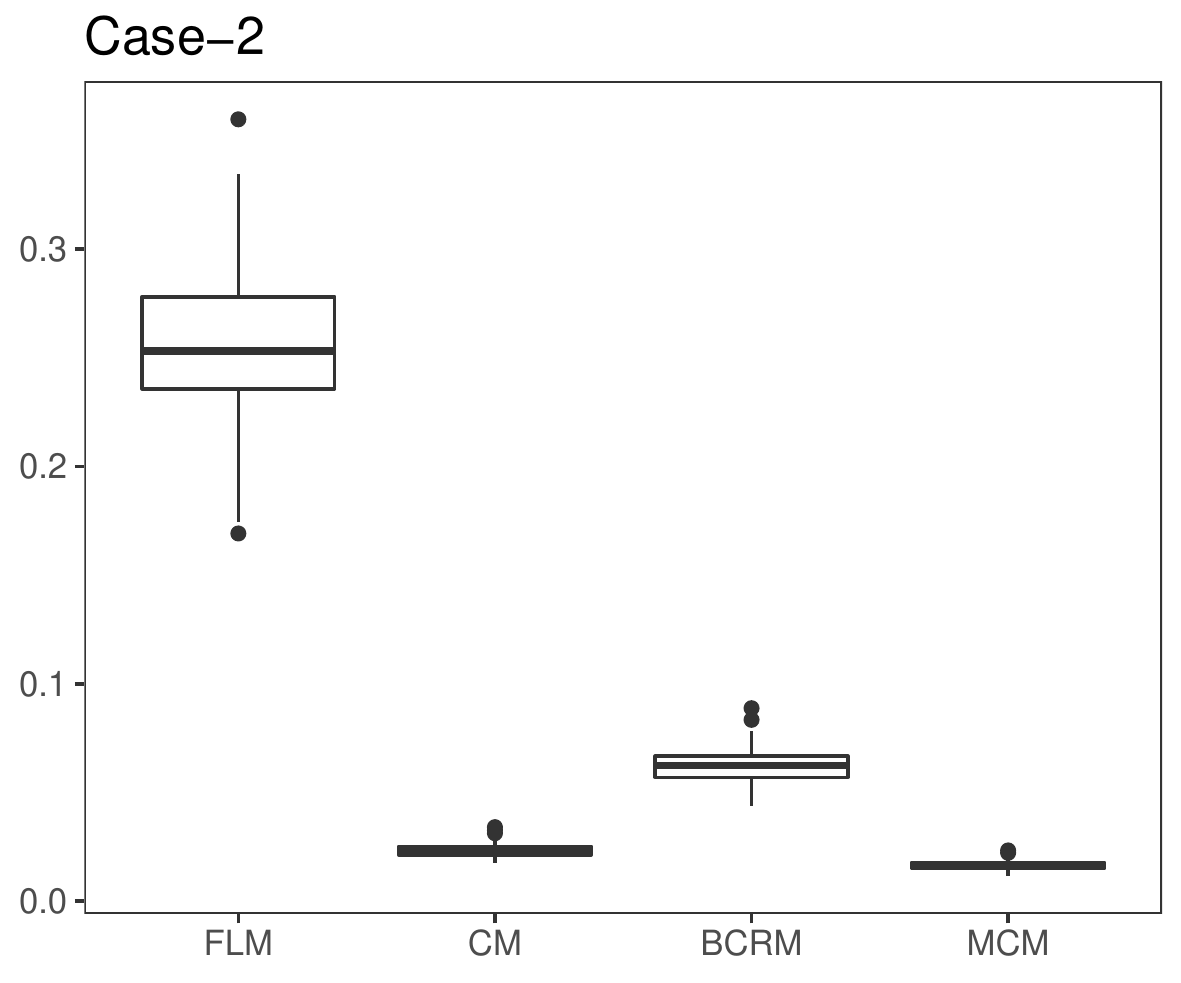}
\\  
  \includegraphics[width=8.5cm]{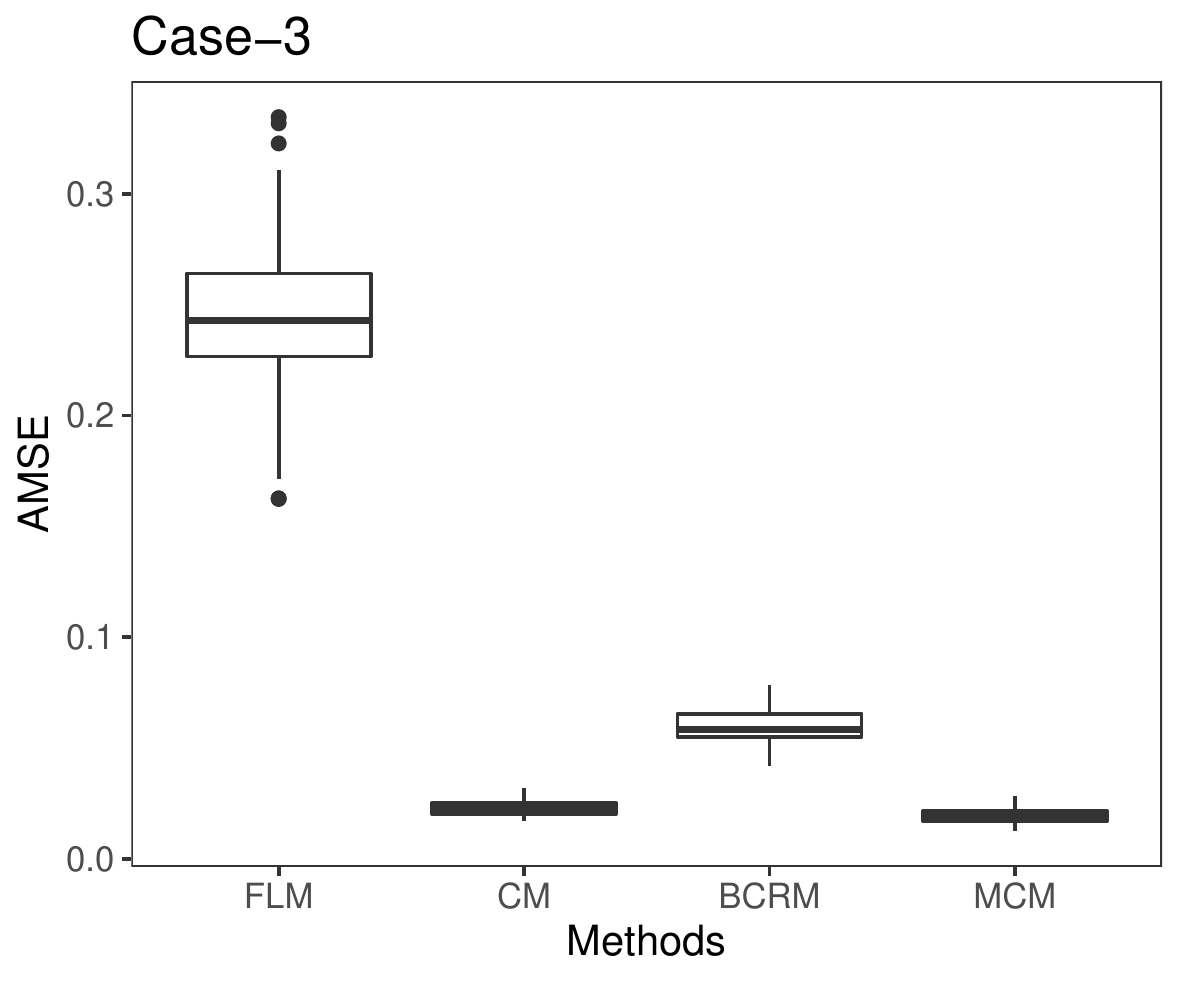}
\qquad
  \includegraphics[width=8.5cm]{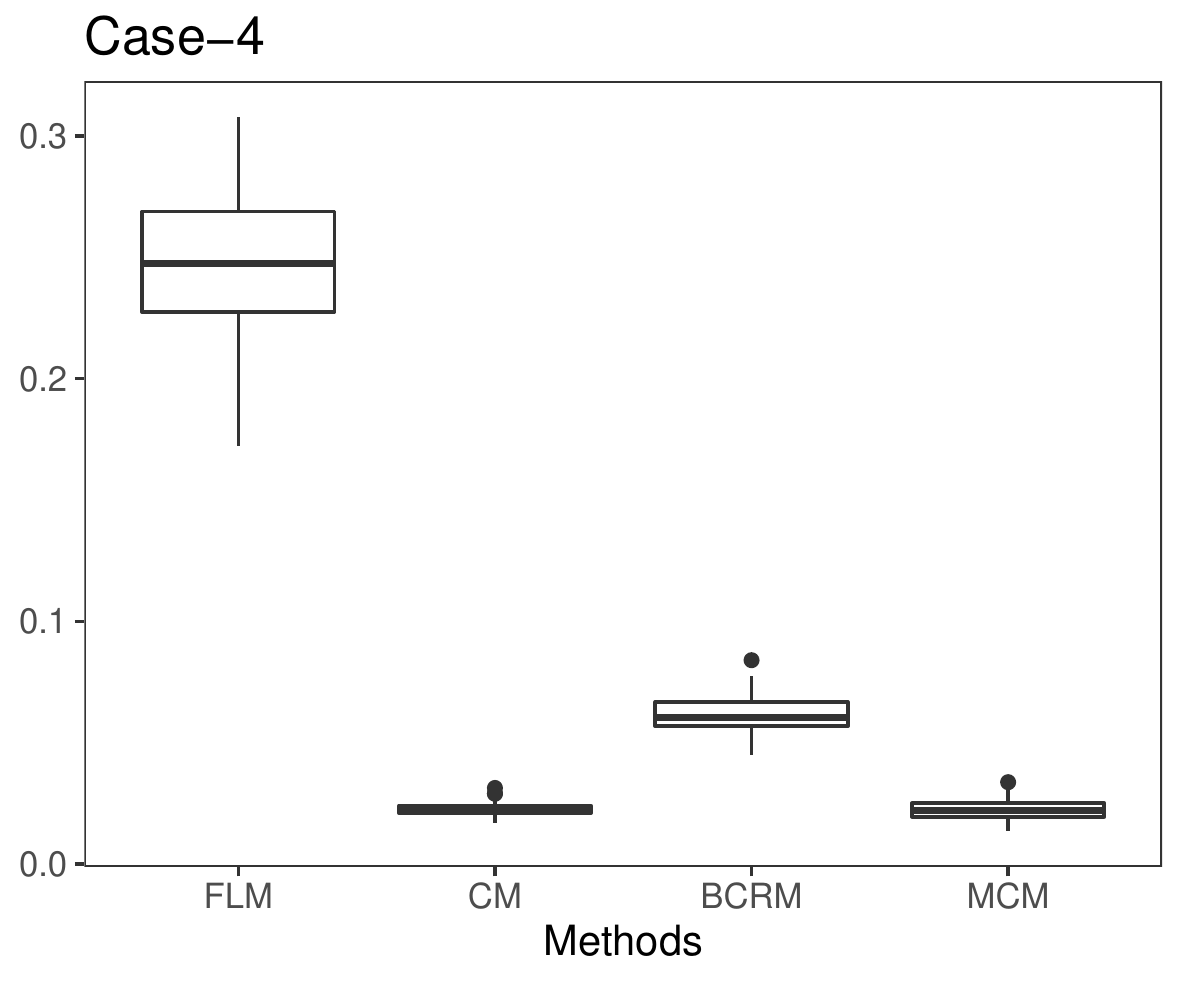}
  \caption{Calculated AMSE values of the FLM, CM, BCRM, and MCM for the lower limit function of the response variable in all cases: Case-1: $(a,b) = (1, 1.5)$, $(c,d) = (1, 1.5)$, Case-2: $(a,b) = (1, 3)$, $(c,d) = (1, 3)$, Case-3: $(a,b) = (3, 5)$, $(c,d) = (5, 8)$, and Case-4: $(a,b) = (8, 20)$, $(c,d) = (6, 15)$.}
  \label{fig:siml}
\end{figure}

\begin{figure}[!t]
  \centering
  \includegraphics[width=8.5cm]{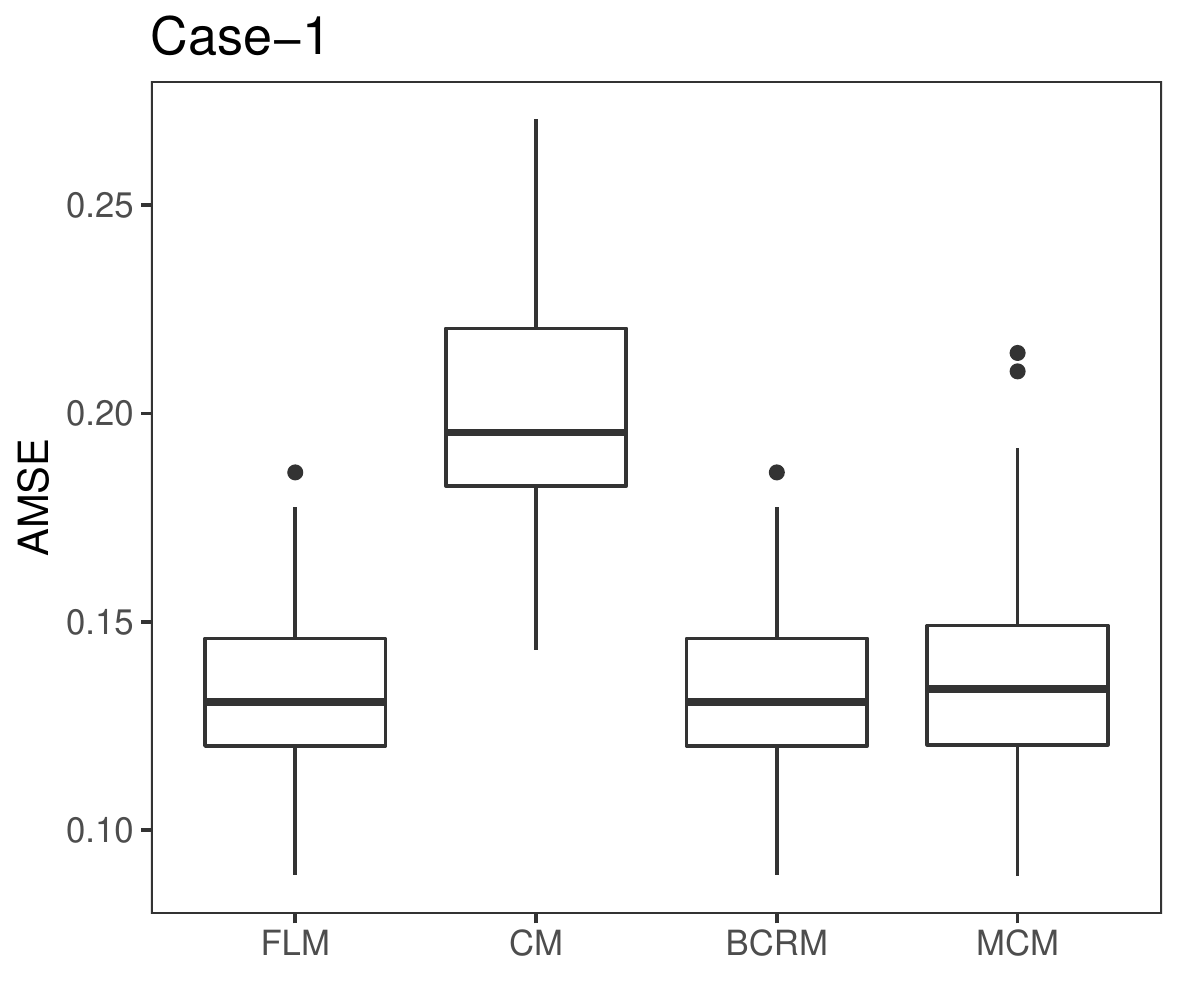}
\qquad
  \includegraphics[width=8.5cm]{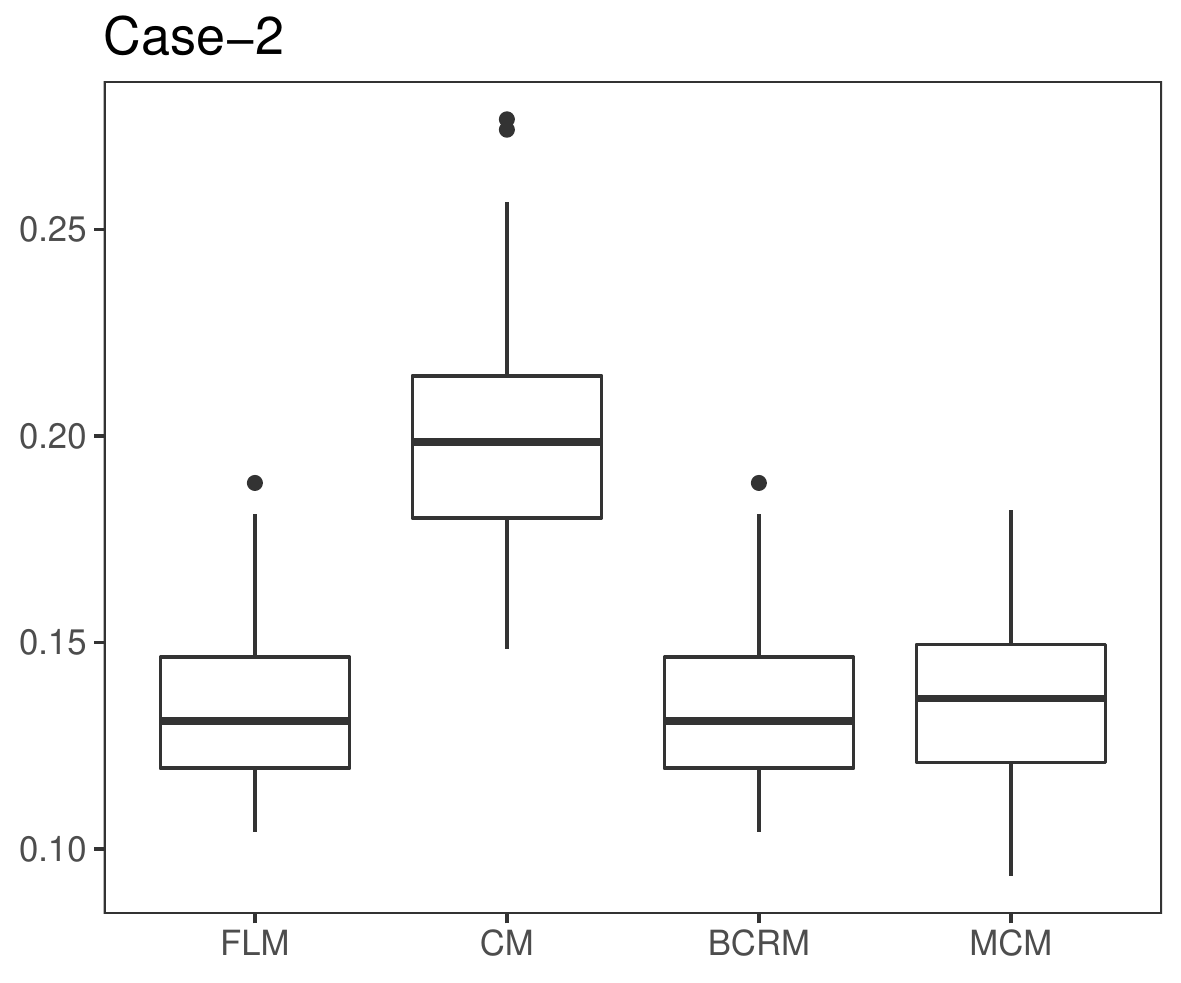}
\\  
  \includegraphics[width=8.5cm]{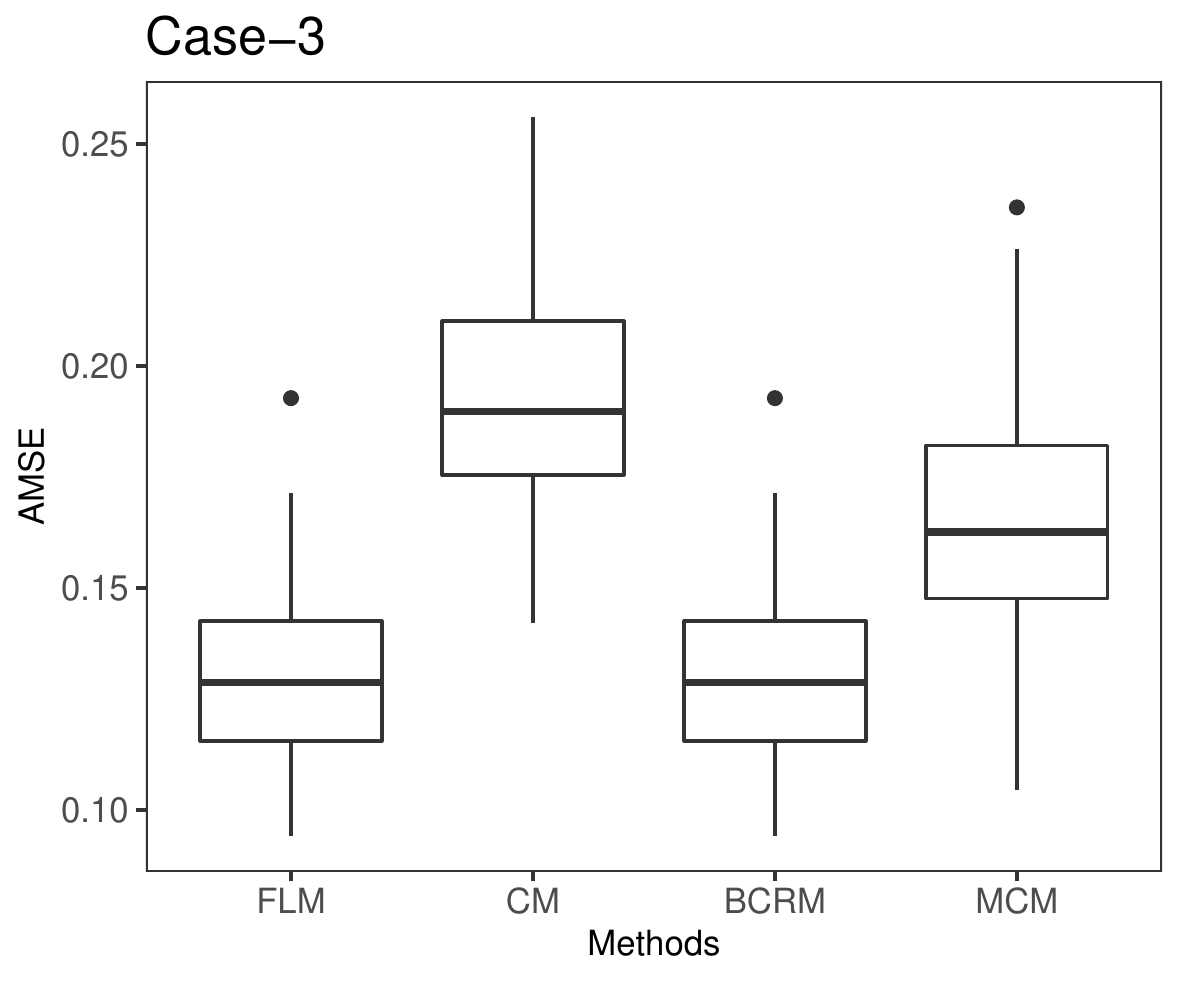}
\qquad
  \includegraphics[width=8.5cm]{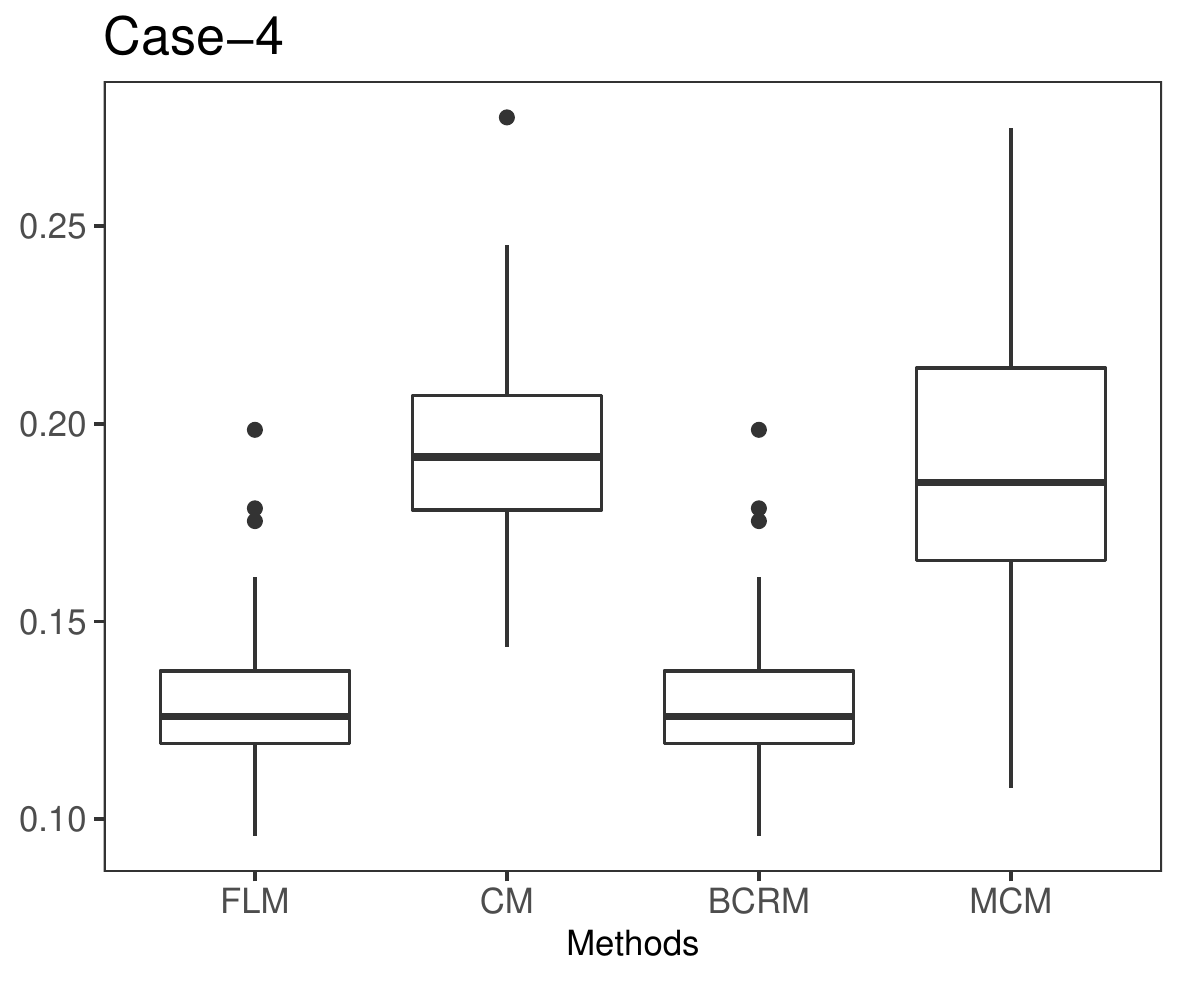}
  \caption{Calculated AMSE values of the FLM, CM, BCRM, and MCM for the upper limit function of the response variable in all cases: Case-1: $(a,b) = (1, 1.5)$, $(c,d) = (1, 1.5)$, Case-2: $(a,b) = (1, 3)$, $(c,d) = (1, 3)$, Case-3: $(a,b) = (3, 5)$, $(c,d) = (5, 8)$, and Case-4: $(a,b) = (8, 20)$, $(c,d) = (6, 15)$.}
  \label{fig:simu}
\end{figure}

Our findings showed that functional CRM and BCRM produce very close results for each other, and thus, we presented the results of the functional BCRM only. The results for the lower and upper limit functions are presented in Figures~\ref{fig:siml} and~\ref{fig:simu}, respectively. The results demonstrate that, for the lower limit functions, the proposed interval-valued functional regression models outperform the FLM for all cases. Compared with functional BCRM, the functional CM and MCM produce better AMSE values (see Figure~\ref{fig:siml}). On the other hand, for the upper limit functions, the functional CM performs less among others for all cases. The FLM, BCRM, and MCM tend to produce similar AMSE values when the range between the lower and upper limit functions is small (Case-1 and Case-2) while the FLM and BCRM produce better results compared to the MCM as the range increases (Case-2 and Case-3). The results presented in Figures~\ref{fig:siml} and~\ref{fig:simu} also show that the FLM and proposed interval-valued functional regression models (except functional MCM) are not affected by the range when predicting lower and upper limit functions. Only the performance of the functional MCM gets worse as the range increases when predicting upper limit functions. 

\begin{figure}[!htbp]
  \centering
  \includegraphics[width=8.5cm]{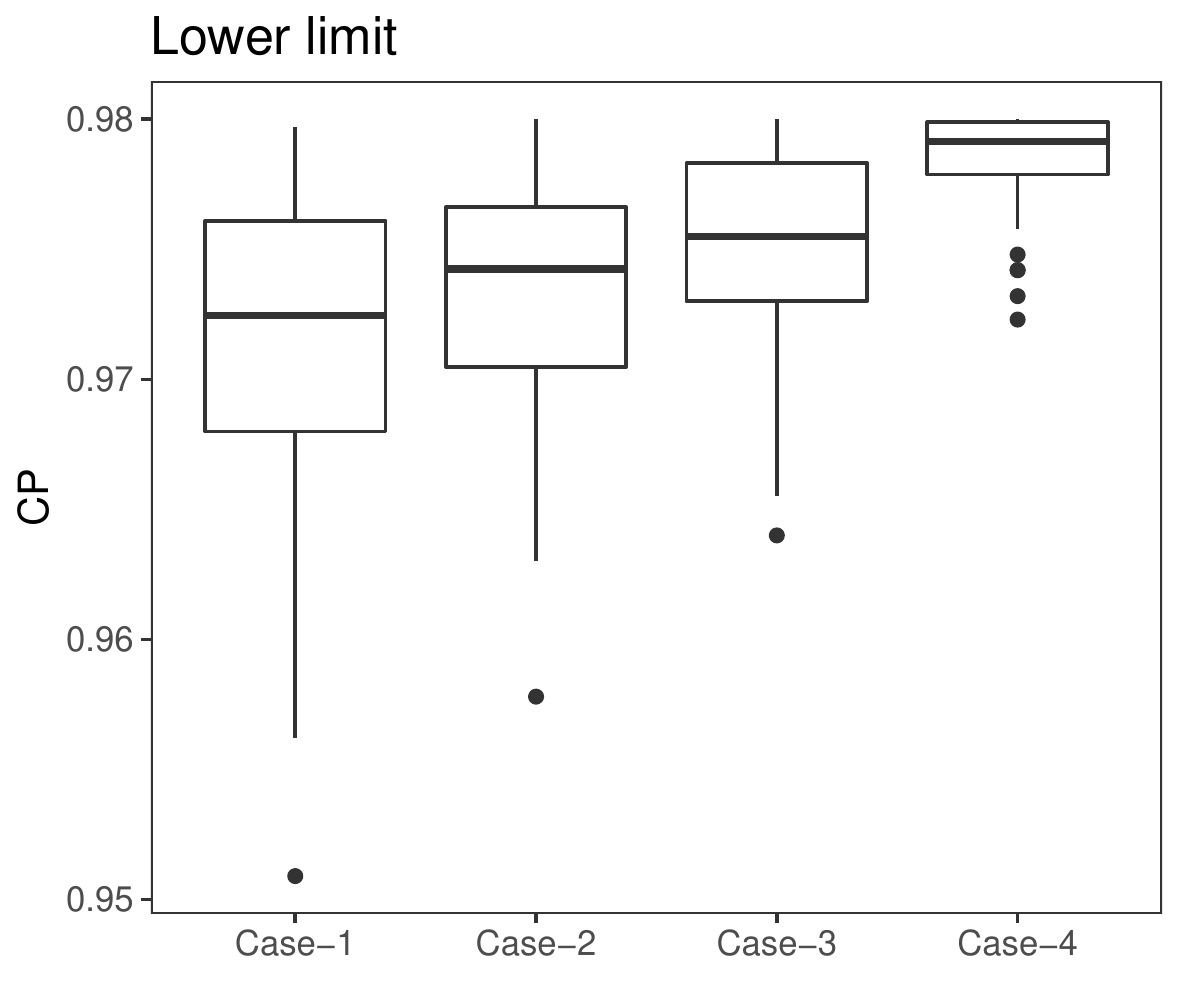}
\qquad
  \includegraphics[width=8.5cm]{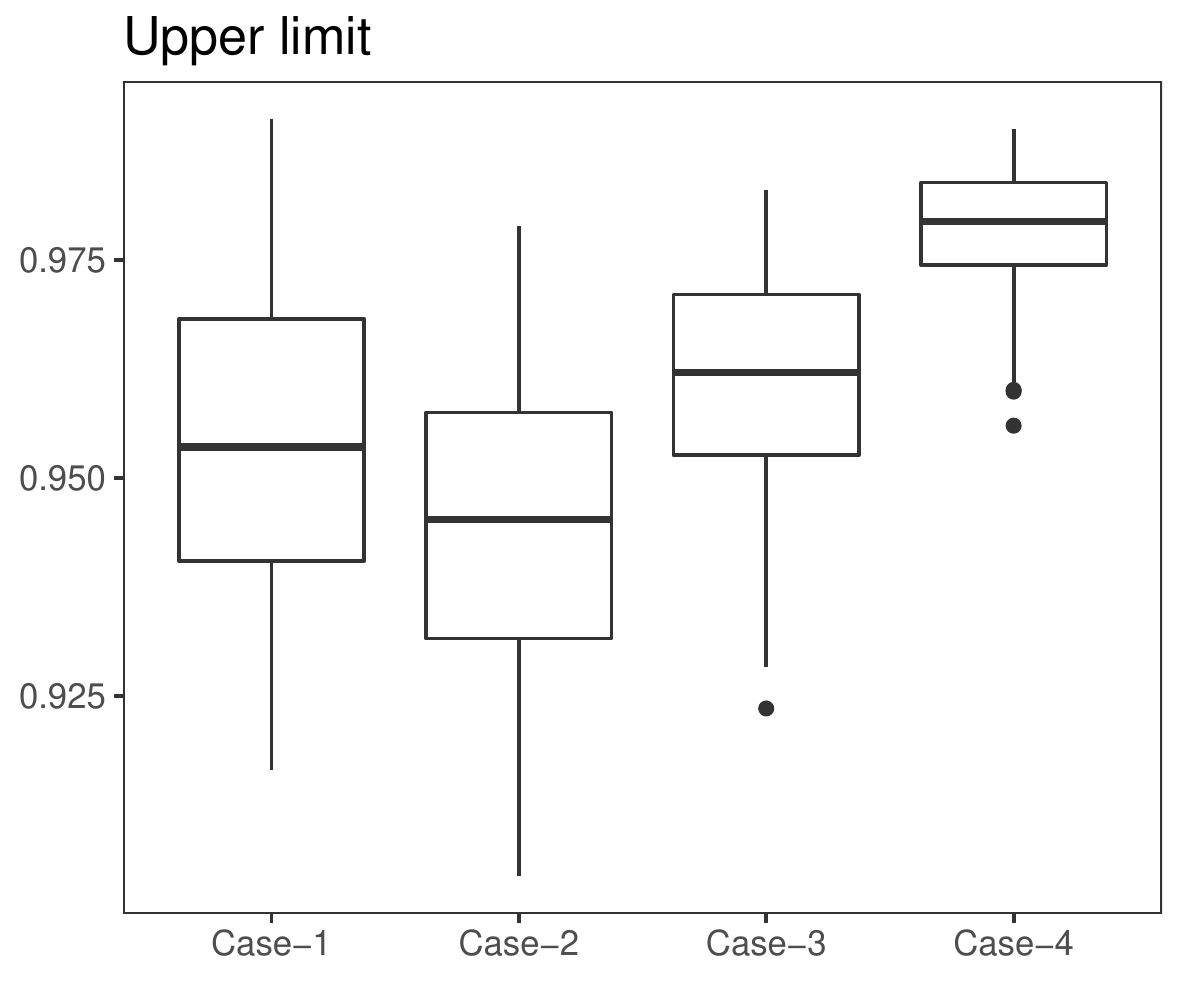}
  \caption{Calculated coverage probabilities of the MCM-based prediction intervals for the lower and upper limit functions in all cases: Case-1: $(a,b) = (1, 1.5)$, $(c,d) = (1, 1.5)$, Case-2: $(a,b) = (1, 3)$, $(c,d) = (1, 3)$, Case-3: $(a,b) = (3, 5)$, $(c,d) = (5, 8)$, and Case-4: $(a,b) = (8, 20)$, $(c,d) = (6, 15)$.}
  \label{fig:simcp}
\end{figure}

The coverage performances of the functional MCM-based prediction intervals are presented in Figure~\ref{fig:simcp}. It is clear from this figure that the functional MCM is capable of producing valid prediction intervals for upper limit functions and generally produces coverage probabilities close to the nominal significance level. On the other hand, for the lower limit functions, over-coverage is observed for all cases.

\subsection{Empirical data example: Oman weather data}\label{sec:real}

\begin{figure}[!htbp]
  \centering
  \includegraphics[width=5.9cm]{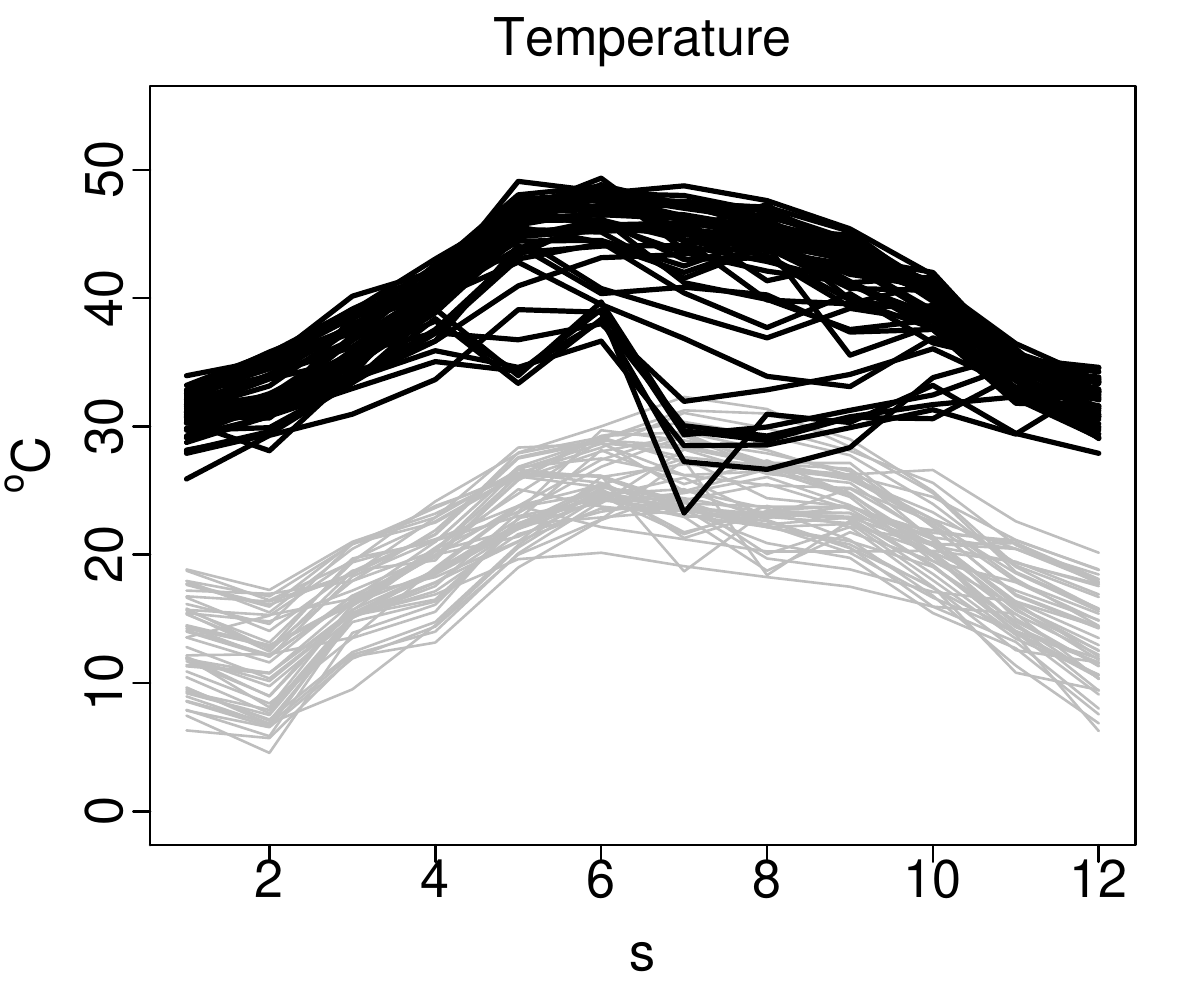}
  \includegraphics[width=5.9cm]{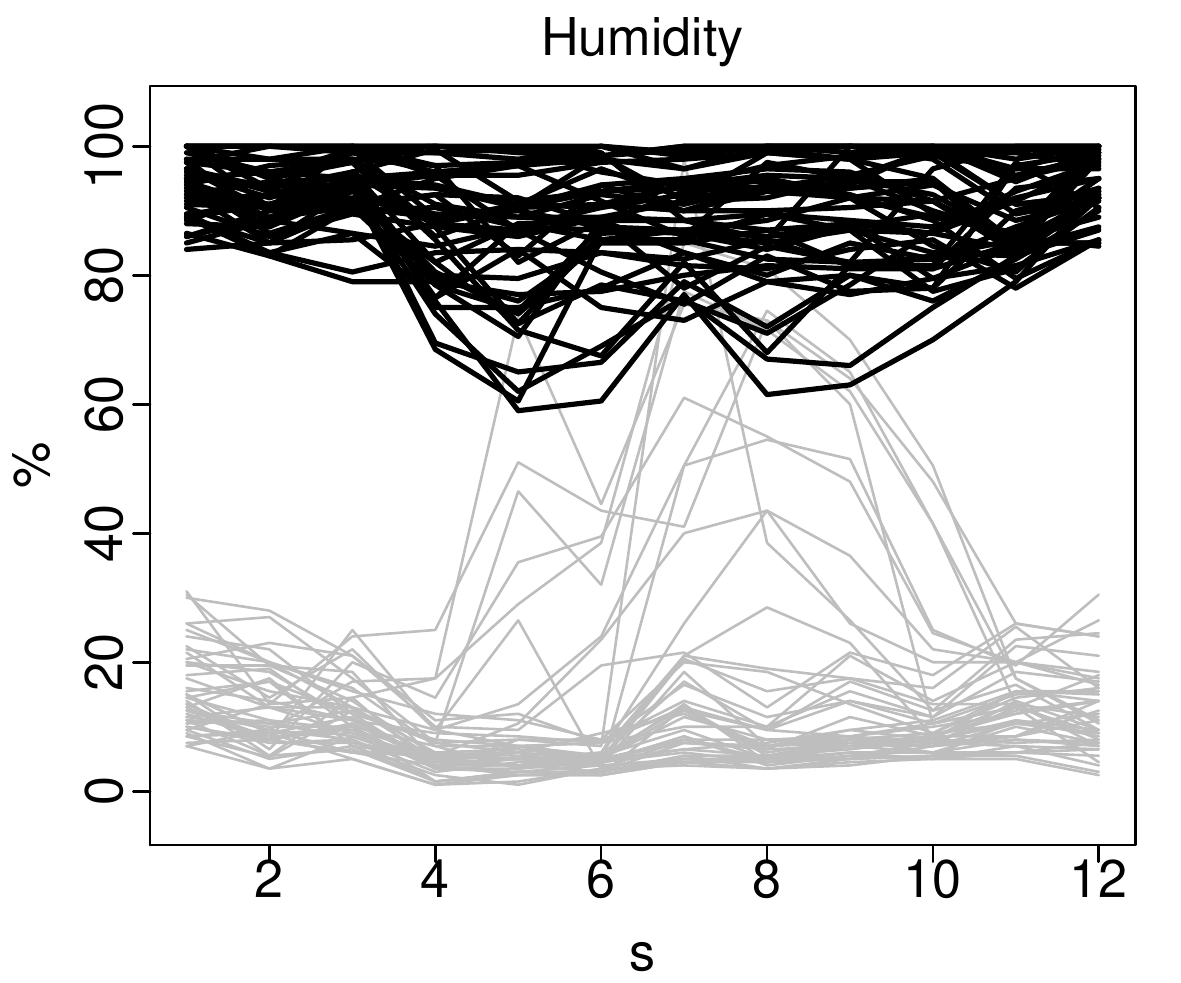}
  \includegraphics[width=5.9cm]{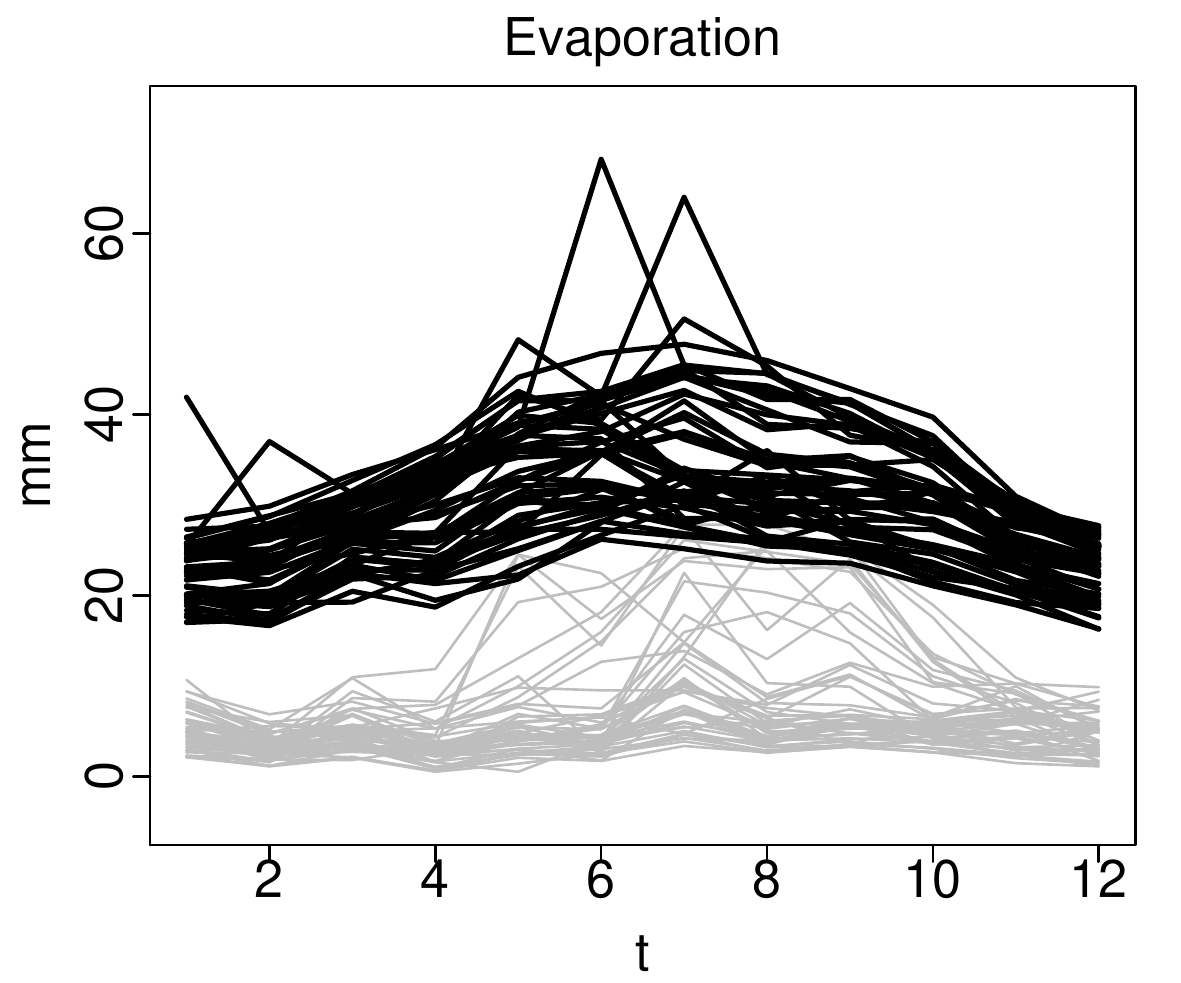}
  \caption{Time series plots of the averaged minimum (gray lines) and maximum (black lines) monthly Oman weather variables.}
  \label{fig:tsplots}
\end{figure}

The empirical data example represents the monthly Oman weather data spanning from January 2017 to December 2018. The dataset contain three interval-valued variables: monthly evaporation (mm), humidity (\%), and temperature ($^\circ$C), and were collected from 48 stations across Oman (dataset is available from the National Center for Statistics \& Information: \url{https://data.gov.om}). The list of stations is reported in Table~\ref{tab:stations}. The data were averaged for each interval-valued variable over the whole time, and the observations were considered as the functions of months, $1 \leq s,t \leq 12$. The plots of the averaged interval-valued functional variables for all 48 stations are presented in Figure~\ref{fig:tsplots}. 

\begin{table}[htbp] \small
\centering
\tabcolsep 0.2in
\caption{Station names for the Oman monthly weather data.}
\begin{tabular}{@{}llllll@{}}
\toprule
Station & Station & Station & Station & Station  \\
\midrule
Adam Airport	&Bidiyah 		& Khasab Port		& Muscat City  		& Shalim  \\
AIJubah			&Bowsher 	 	& Liwa 				& Qairoon Hairiti 	& Sohar Airport \\
Al Amrat		&Bukha 	 		& Madha 			& Qalhat  			& Sunaynah \\
Al Hamra		&Dhank Qumaira	& Mahdah 			& Qarn alam 		& Sur  \\
Al Jazir		&Diba 			& Majis 			& Qurayyat 			& Suwaiq  \\
Al Mazyunah		&Duqum Airport	& Masirah 			& Ras AlHaad  		& Taqah  \\
Al Mudhaibi 	&Ibra 			& Mina Salalah		& Rustaq   			& Thamrayt  \\
Al-Buraymi		&Ibri 			& Mina Sultan Qaboos& Sadah   			& Yanqul \\
Bahla 			&Izki 	 		& Mirbat 			& Salalah Airport   \\
Bidbid 			&Khasab Airport	& Muqshin 			& Samail   \\
\bottomrule
\end{tabular}
\label{tab:stations}
\end{table}

\begin{figure}[!htbp]
  \centering
  \includegraphics[width=8.5cm]{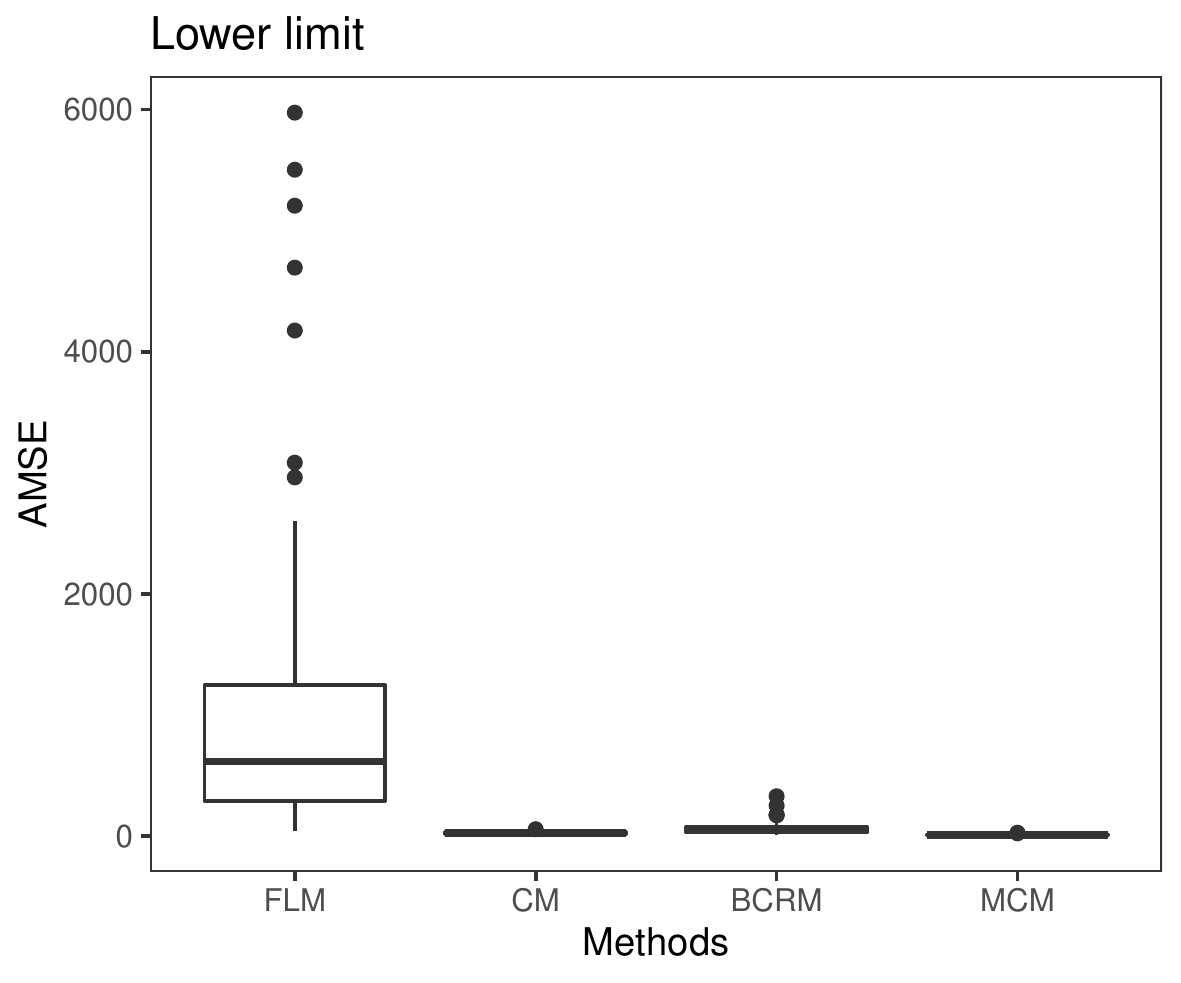}
\qquad
  \includegraphics[width=8.5cm]{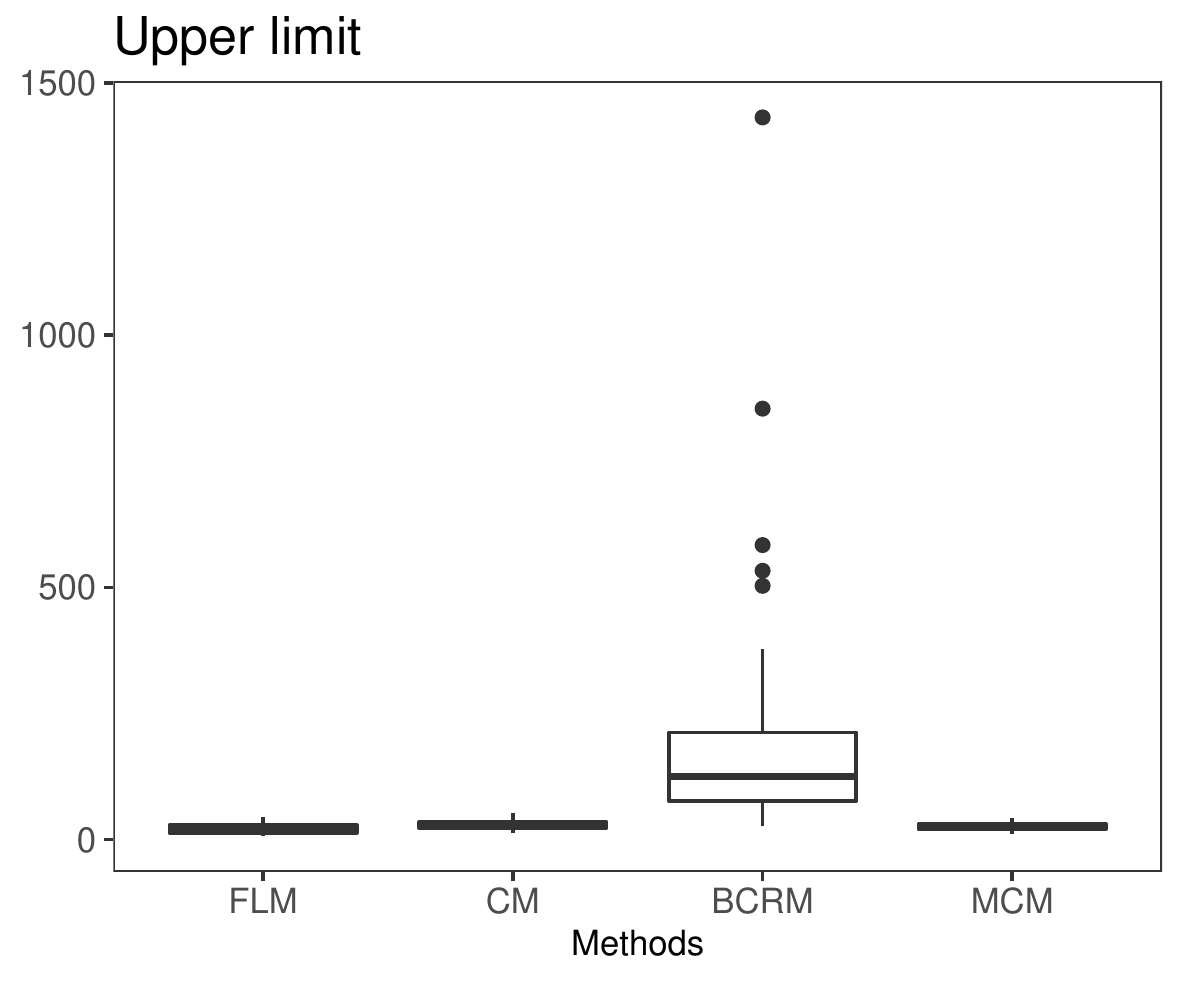}
  \caption{Estimated AMSE values for the monthly Oman weather data.}
  \label{fig:amse_real}
\end{figure}

We considered predicting monthly minimum and maximum evaporation using the minimum and maximum temperature and humidity variables. For this purpose, the values of the discretely observed minimum and maximum variables (and their center and range points) were first converted to functional forms by B-spline basis function expansion using 10 number of basis functions. The following procedure was repeated 100 times to evaluate the predictive performance of the proposed interval-valued functional regression models as well as the traditional FLM. In each repeat, the dataset was divided into two parts; the models were constructed based on the functions of 40 randomly selected stations to predict the minimum and maximum evaporations of the remaining eight stations. For each model, the parameter matrix $\mathbf{B}$ was estimated using the ML method, as explained in Section~\ref{sec:methodology}. For each replicate, the AMSE$^l$ and AMSE$^u$ values were calculated. Our findings are presented in Figure~\ref{fig:amse_real}.

\begin{figure}[!htbp]
  \centering
  \includegraphics[width=10cm]{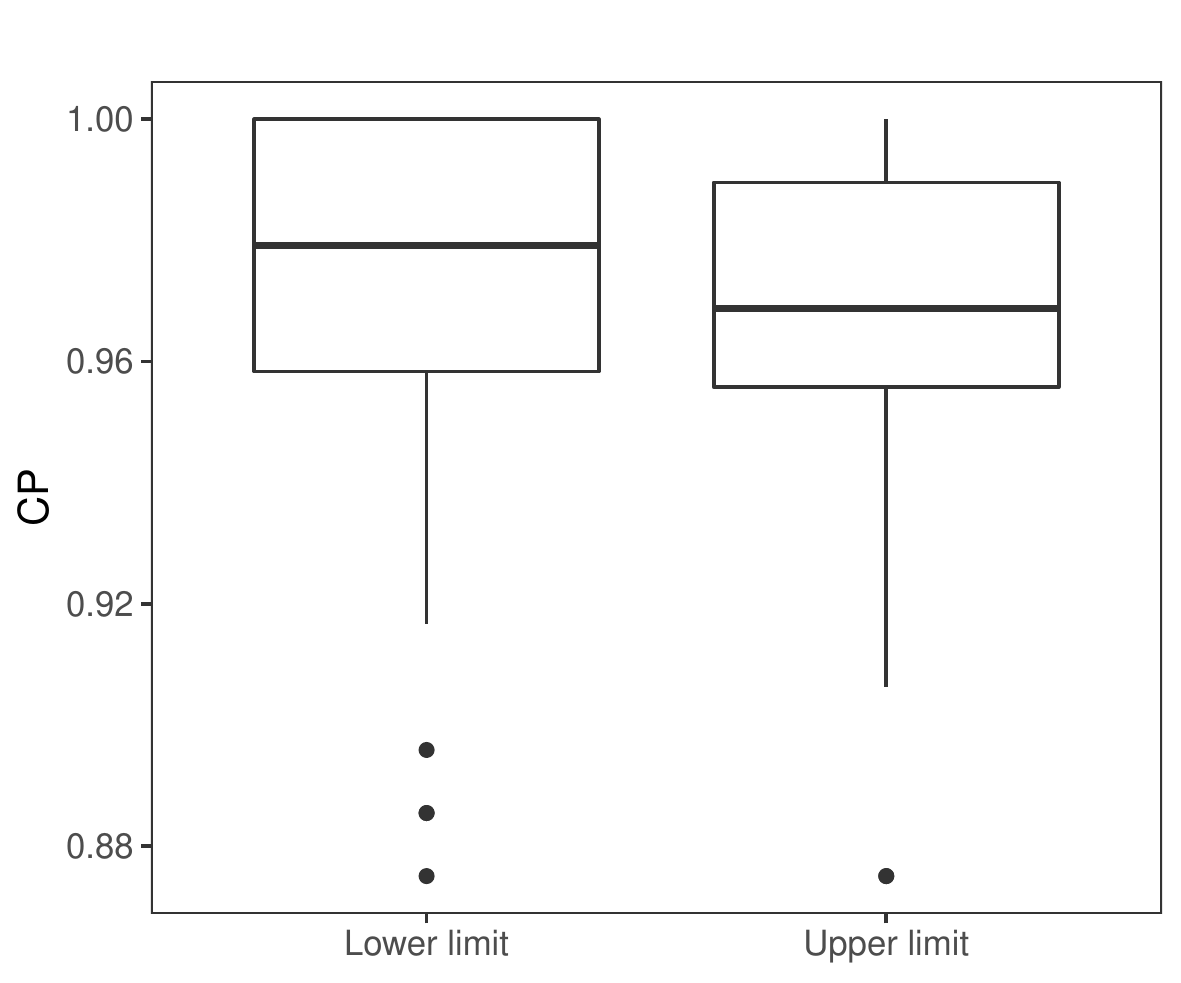}
  \caption{Estimated AMSE values for the monthly Oman weather data.}\label{fig:ci_real}
\end{figure}

The results show that all the proposed methods outperform the FLM for the lower limit functions. The functional CM and MCM produce slightly less AMSE$^l$ compared with BCRM. On the other hand, for the upper limit functions, the functional BCRM performs less, among others. The functional MCM and FLM produce slightly better performances compared with the functional CM. Also, the functional MCM-based prediction intervals were calculated, and the results are given in Figure~\ref{fig:ci_real}. This figure shows that the functional MCM generally produce valid prediction intervals for both lower and upper limit functions. However, as it is seen from Figure~\ref{fig:ci_real}, MCM produces low coverage probabilities for some cases. Because the dataset includes some outlying observations, the performance of MCM is affected by the presence of outlying observations. A robust estimation method may be used to overcome this problem. 

\begin{figure}[!htbp]
  \centering
  \includegraphics[width=12.1cm]{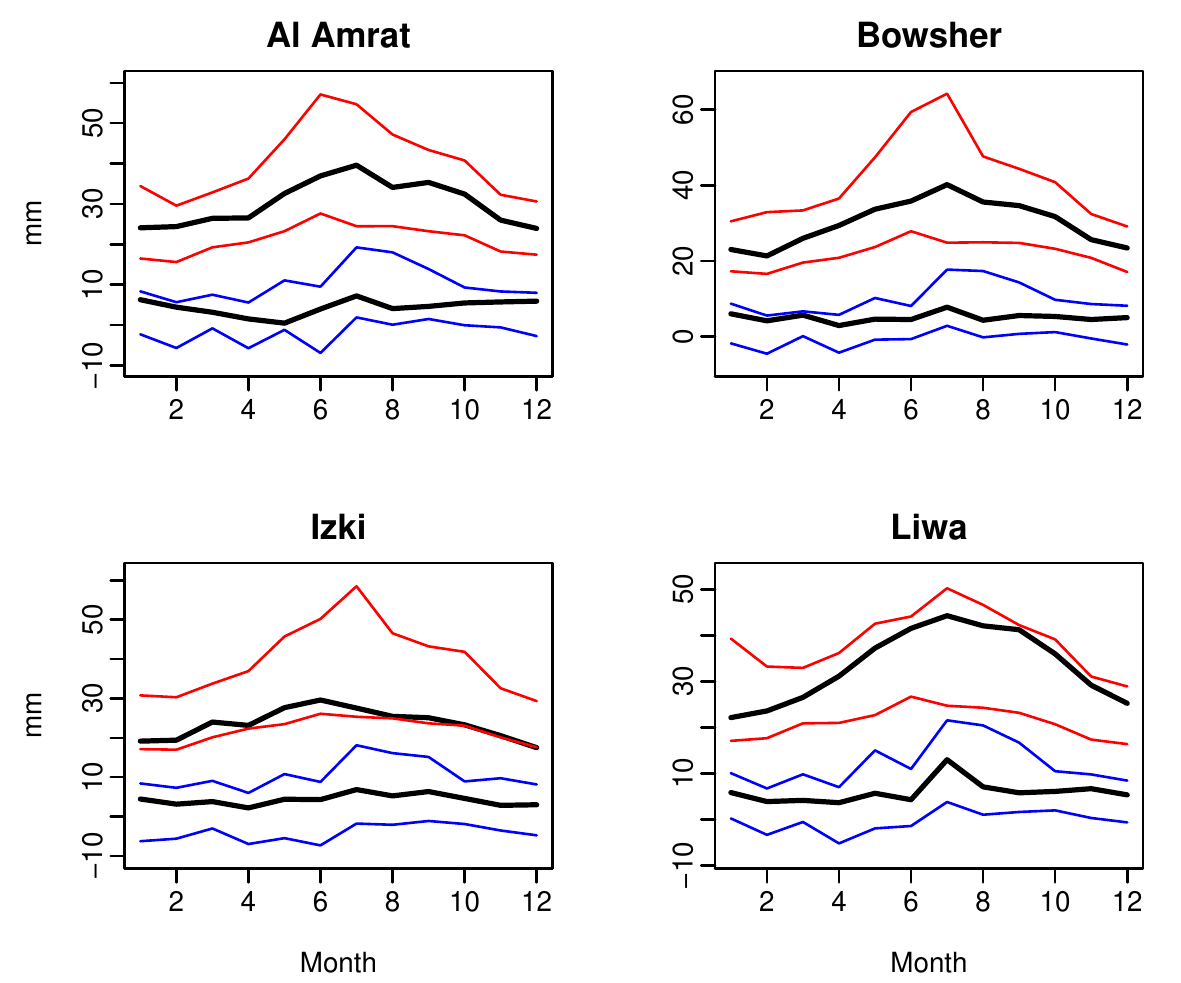}
  \caption{Plots of the observed lower and upper limit functions (black lines) and their calculated MCM-based 95\% prediction intervals for four stations; Al Amrat, Bowsher, Izki, and Liwa. Prediction intervals for the lower and upper limit functions are given in blue and red colors, respectively.}
  \label{fig:ci_ex}
\end{figure}

An example of the constructed MCM-based prediction intervals is presented in Figure~\ref{fig:ci_ex}.

\section{Conclusion}\label{sec:conc}

Interval-valued data regression models have become a natural framework to analyze datasets that are collected in an interval. On the other hand, recent technological advances in data collection tools cause complex and high dimensional datasets, which may not be analyzed by existing traditional methods. FDA tools are one of the frequently used methods for visualizing and analyzing such data. Particularly, the functional linear models have been one of the most commonly used techniques to explore the association between the functional response and predictor variables. Although the functional extensions of many traditional methods are available in the literature, the functional forms of the interval-valued data regression models have not yet been studied.

In this study, we present the functional forms of some well known interval-valued data regression models and examine their prediction performances. The proposed methods are based on the function-on-function regression model, where both the response and predictors are functions which is common in real-life situations. The finite sample performance of the proposed interval-valued functional regression models is evaluated via Monte Carlo simulations and an empirical data analysis. Also, we compare the finite sample performance of the proposed methods with the traditional functional linear model. Our findings show that the proposed regression models are superior to the traditional functional linear model when predicting lower limit functions. Also, they produce competitive performance with FLM when predicting upper limit functions. Moreover, the proposed functional MCM is used to construct prediction intervals for both lower and upper limit functions of the response variable. All the numerical analyses considered in this study have shown that the functional MCM is generally capable of producing valid prediction intervals.

For future studies, the followings may be considered;
\begin{inparaenum}
\item[1)] throughout this study, only the function-on-function regression is considered, but the proposed methods can also be extended to other functional regression models such as scalar-on-function and function-on-scalar,
\item[2)] we consider only the B-spline basis function to convert discretely observed data to functional form, other basis functions, such as Fourier, wavelet, and radial basis functions may also be considered, and 
\item[3)] the model parameters of the proposed interval-valued functional regression models are estimated using the ML method, other estimation methods, such as partial least squares and principal component regression, may also be used.
\end{inparaenum}

\end{document}